\documentclass[final,5p,times,twocolumn]{elsarticle}

\usepackage{graphics}
\usepackage{graphicx}

\usepackage{amssymb}

\usepackage{algorithm}
\usepackage{algorithmic}
\usepackage{url}
\usepackage{booktabs}
\usepackage{wrapfig}
\usepackage{listings}
\usepackage{amsthm}
\usepackage{array}
\usepackage{pbox}
\usepackage{graphicx}
\usepackage{multirow}
\usepackage{adjustbox}
\usepackage{lscape}
\usepackage{url}
\usepackage{amssymb}
\usepackage{soul} 

\usepackage{color}
\newcommand\eat[1]{}

\newcommand{\mr}[2][c]{\begin{tabular}[#1]{@{}c@{}}#2\end{tabular}}

\graphicspath{{images/}}

\journal{ }

\begin{document}

\begin{frontmatter}



\title{The Many Faces of Data-centric Workflow Optimization: A Survey}


\author{Georgia Kougka}
\ead{georkoug@csd.auth.gr}
\address{Aristotle University of Thessaloniki, Greece}

\author{Anastasios Gounaris}
\ead{gounaria@csd.auth.gr}
\address{Aristotle University of Thessaloniki, Greece}

\author{Alkis Simitsis}
\ead{alkis@hpe.com}
\address{HP Labs, Palo Alto, USA}

\begin{abstract}
Workflow technology is rapidly evolving and, rather than being limited to modeling the control flow in business processes, is becoming a key mechanism to perform advanced data management, such as big data analytics. This survey focuses on data-centric workflows (or workflows for data analytics or data flows), where a key aspect is data passing through and getting manipulated by a sequence of steps. The large volume and variety of data, the complexity of operations performed, and the long time such workflows take to compute give rise to the need for optimization. In general, data-centric workflow optimization is a technology in evolution. This survey focuses on techniques applicable to workflows comprising arbitrary types of data manipulation steps and semantic inter-dependencies between such steps. Further, it serves a twofold purpose. Firstly, to present the main dimensions of the relevant optimization problems and the types of optimizations that occur before flow execution. Secondly, to provide a concise overview of the existing approaches with a view to highlighting key observations and areas deserving more attention from the community.
\end{abstract}


\end{frontmatter}


\section{Introduction}
\label{sec:intro}
Workflows aim to model and execute  real-world intertwined or interconnected processes, named as \emph{tasks} or  \emph{activities}. While this is still the case, workflows play an increasingly significant role in processing very large volumes of data, possibly under highly demanding requirements.
Scientific workflow systems tailored to data-intensive e-science applications have been around since the last decade, e.g., \cite{CG08,DGST09}. This trend is nowadays complemented by
the evolution of workflow technology to serve (big) data analysis, in settings such as business intelligence, e.g., \cite{CDN11}, and business process management, e.g., \cite{BHS09}. Additionally, massively parallel  engines, such as Spark, are becoming increasingly popular for designing and executing workflows.

Broadly, there are two big workflow categories, namely \emph{control-centric} and \emph{data-centric}. A workflow is commonly represented as a directed graph, where each task corresponds to a node in the graph and the edges represent the \emph{control flow} or the \emph{data flow}, respectively. The \emph{control-centric workflows} are most often encountered in business process management \cite{BS14} and they emphasize the passing of control across tasks and gateway semantics, such as branching execution, iterations, and so on; transmitting and sharing data across tasks is a second class citizen. In control-centric workflows, only a subset of the graph nodes correspond to activities, while the remainder denote events and gateways, as in the BPMN standard. In
\emph{data-centric workflows} (or workflows for data analytics or simply \emph{data flows}\footnote{Hereafter, these three terms will be used interchangeably; the terms workflow and flow will be used interchangeably, too.}), the graph is typically acyclic (directed acyclic graph - DAG). The nodes of the DAG represent solely actions related to the manipulation, transformation, access and storage of data,
e.g., as in \cite{DCSW09,OVDP11,SWCD12,ZBML09} and in popular data flow systems, such as Pentaho Data Integration (Kettle) and Spark.
The tokens passing through the tasks correspond to processed data. The control is modeled implicitly assuming that each task may start executing when the entire or part of the input becomes available. 
This survey considers data-centric flows exclusively.

Executing data-centric flows efficiently is a far from trivial issue. Even in the most widely used data flow tools, flows are commonly designed manually. Problems in the optimality of those designs stem from the complexity of such flows and the fact that in some applications, flow designers might not be systems experts \cite{beckman13} and consequently, they tend to design with only semantic correctness in mind. In addition, executing flows in a dynamic environment may entail that an optimized design in the past may behave suboptimally in the future due to changing conditions \cite{HDP14,SWDH13}.

The issues above call for a paradigm shift in the way data flow management systems are engineered and more specifically, there is a growing demand for automated optimization of flows. An analogy with database query processing, where declarative statements, e.g., in SQL, are automatically parsed, optimized, and then passed on to the execution engine is drawn. But data flow optimization is more complex, because tasks need not belong to a predefined set of algebraic operators with clear semantics and there may be arbitrary dependencies among their execution order. In addition, in data flows there may be optimization criteria apart from performance, such as reliability and freshness depending on business objectives and execution environments~\cite{SWCD09}. This survey covers optimization techniques\footnote{The terms \emph{technique, proposal}, and \emph{work} will be used interchangeably.} applicable to data flows, including database query optimization techniques that consider arbitrary plan operators, e.g., user-defined functions (UDFs), and dependencies between them. To the contrary, we do not aim to cover techniques that perform optimizations considering solely specific types of tasks, such as filters, joins and so on.

The contribution of this survey is the provision of a taxonomy of data flow optimization techniques that refer to the flow plan generation layer. In addition, a concise overview of the existing approaches with a view to (i) explaining the technical details and the distinct features of each approach in a way that facilitates result synthesis; and (ii) highlighting strengths and weaknesses, and areas deserving more attention from the community is provided.

The main findings are that on the one hand, big advances have been made and most of the aspects of data flow optimization have started to be investigated. On the other hand, data flow  optimization is rather a technology in evolution. Contrary to query optimization, research so far seems to be less systematic and mainly consists of  ad-hoc techniques, the combination of which is unclear.

The structure of the rest of this article is as follows. The next section describes the survey methodology and provides details about the exact context considered. Section~\ref{sec:taxonomy} presents a taxonomy of existing optimizations that take place before the flow enactment. Section~\ref{sec:classification} describes the state-of-the-art techniques grouped by the main optimization mechanism they employ. Section~\ref{sec:eval} presents the ways in which  optimization proposals for data-centric workflows have been evaluated. Section~\ref{sec:discussion} highlights our findings. Section~\ref{sec:other} touches upon tangential flow optimization-related techniques that have recently been developed along with scheduling optimizations taking place during flow execution.
Section~\ref{sec:rw} reviews surveys that have been conducted in related areas and finally, Section~\ref{sec:summary} concludes the paper.

\section{Survey Methodology}
\label{sec:methodology}

We first detail our context with regards to the architecture of a Workflow Management System (WfMS). Then we explain the methodology for choosing the techniques included in the survey and their dimensions, on which we focus. Finally, we summarize the survey contributions.

\subsection{Our Context within WfMSs}
The life cycle of a workflow can be regarded as  an iteration of four phases, which cover every stage from the workflow modeling until its output analysis \cite{LPVM15}. The four phases are \emph{composition, deployment, execution}, and \emph{analysis} \cite{LPVM15}. The type of workflow optimization, on which this work focuses, is part of the deployment phase where the concrete executable workflow plan is constructed defining execution details, such as the engine that will execute each task. Additionally, Liu et al. \cite{LPVM15} introduce a functional architecture for each data-centric Workflow Management System (WfMS), which consists of five layers: i) \emph{presentation}, which comprises the user interface; ii) \emph{user services}, such as the workflow monitoring and data provision components; iii) \emph{workflow execution plan (WEP) generation}, where the workflow plan is optimized, e.g., through workflow refactoring and parallelization, and the details needed by the execution engine are defined; iv) \emph{WEP execution}, which deals with the scheduling and execution of the (possibly optimized) workflow, but also considers fault-tolerance issues, and finally, v) the \emph{infrastructure} layer, which provides the interface between the workflow execution engine and the underlying physical resources.

According to the above architecture, one of the roles of a WfMS is to compile and optimize the workflow execution plans just before the workflow execution. Optimization of data flows, as conceived in this work, forms an essential part of the WEP generation layer and not of the execution layer. Although there might be optimizations in the WEP execution layer as well, e.g., while scheduling the WEP, these are out of our scope. More specifically, the mapping of flow tasks to concrete processing nodes during execution, e.g, task $X$ of the flow should run on processing node $Y$,  is traditionally considered to be a scheduling activity that is part of WEP execution layer rather than the WEP generation one, on which we focus. Finally, we use the terms task and activity interchangeably, both referring to entities that are not yet instantiated, activated or executed.

\subsection{Techniques Covered}

The main part of this survey covers all the data flow optimization techniques  that  meet the following criteria to the best of authors' knowledge:
\begin{itemize}

\item They refer to the WEP generation layer in the architecture described above.

\item They refer to techniques that are applicable to any type of tasks rather than being tailored to specific types, such as filters and joins.

\item The partial ordering of the flow tasks is subject to dependency (or, else precedence) constraints between tasks, as is the generic case for example of scientific and data analysis flows; these constraints denote whether a specific task must precede another task or not in the flow plan.

\end{itemize}

We surveyed all types of venues where relevant techniques are published. Most of the covered works come from the broader data management and e-science community, but there are proposals from other areas, such as algorithms. We also include techniques that were proposed without generic data flows in mind, but meet our criteria and thus are applicable to generic data flows. An example is the proposal for queries over Web Services (WSs) in \cite{Sriv06}.

\subsection{Technique Dimensions Considered}
\label{sec:dimensions}

We assume that the user initially defines the flow either at a high-level non-executable form or in an executable form that is not optimized. The role of the optimizations considered is to transform the initial flow into an optimized ready-to-be executed one.\footnote{Through considering optimizations starting from a valid initial flow, we exclude from our survey the big area of answering queries in the presence of limited access patterns, in which, the main aim is to construct such an initial plan \cite{Li03,PredaKSNYW10} through selecting an appropriate subset of tasks from a given task pool; however, we have considered works from data integration that optimize the plan after it has been devised, such as \cite{YLUG99} or \cite{Flor99}, which is subsumed by \cite{KougkaG15}.} Analogously to query optimization, it is convenient to distinguish between high-level and low-level flow details. The former capture essential flow parts, such as the final task sequencing, at a higher level than that of complete execution details, whereas the latter include all the information needed for execution. In order to drive the optimization, a set of metadata is assumed to be in place. This metadata can be statistics, e.g., cost per task invocation and size of task output per input data item, information about the dependency constraints between tasks, that is a partial order of tasks, which must be always preserved to ensure semantic correctness, or other types of information as explained in this survey.

To characterize optimizations that take place before the flow execution (or enactment), we pose a set of questions when examining each existing proposal:

\begin{enumerate}

\item \emph{What is the effect on the execution plan?}, which aims to identify the type of incurred enhancements to the initial flow plan.

\item \emph{Why?}, which asks for the objectives of the optimization.

\item \emph{How?}, which aims to clarify the type of the solution.

\item \emph{When?}, to distinguish between cases where the WEP generation phase takes place strictly before the WEP execution one, and where these phases are interleaved.

\item \emph{Where the flow is executed?}, which refers to the execution environment.

\item \emph{What are the requirements?}, which refers to the input flow metadata in order to apply the optimization.

\item \emph{In which application domain?}, which refers to the domain for which the technique initially targets.

\end{enumerate}

We regard each of the above questions as a different dimension. As such, we derive seven dimensions:  (i) the \emph{Mechanisms} referring to the process through which an initial flow is transformed into an optimized one; (ii) the \emph{Objectives} that capture the one or more criteria of the optimization process; (iii) the \emph{Solution Types} defining whether an optimization solution is accurate or approximate with respect to the underlying formulation of the optimization problem; (iv) the \emph{Adaptivity} during the flow execution;  (v) the \emph{Execution Environment} of the flow and its distribution; (vi) the \emph{Metadata} necessary to apply the optimization technique; and finally, (vii) the \emph{Application Domain}, for which each optimization technique is initially proposed.

\section{Taxonomy of Existing Solutions}
\label{sec:taxonomy}

\begin{figure*}[tb!]
\centering
\includegraphics[width=0.7\textwidth]{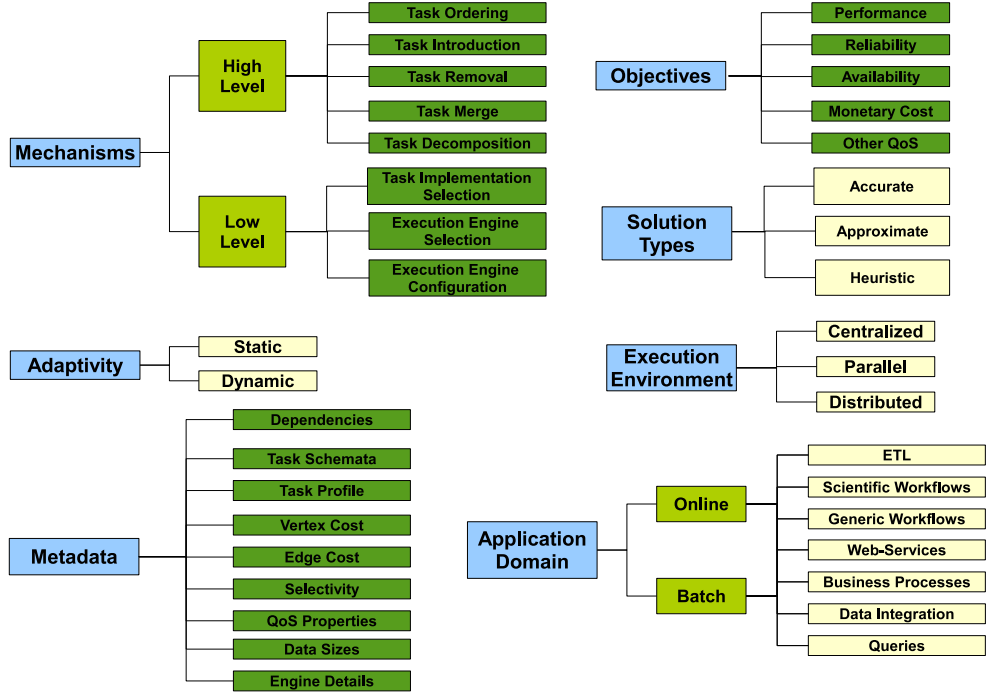}
\caption{A taxonomy of data-centric flow optimization for each of the identified dimensions.}
\label{fig:taxonomy}
\end{figure*}

Based on the dimensions identified above, we build a taxonomy of \emph{existing} solutions. More specifically, for each dimension, we gather the values encountered in the techniques covered hereby.
In other words, the taxonomy is driven by the current state-of-the-art and aims to provide a bird's eye view of today's data flow optimization techniques. The taxonomy is presented in Figure \ref{fig:taxonomy} and analyzed below, followed by a discussion of the main techniques proposed to date in the next section. In the figure, each dimension (in light blue) can take one or more values. Single-value and multi-value dimensions are shown as   yellow and green rectangles, respectively.

\subsection{Flow Optimization Mechanisms}

A data flow is typically represented as a directed acyclic graph (DAG) that is defined as $G=(V,E)$, where $V$ denotes the nodes of the graph corresponding to a set of tasks and $E$ represents a set of pair of nodes, where each pair denotes the data flow between the tasks. If a task outputs data that cannot be directly consumed by a subsequent task, then data transformation needs to take place through a third task; no data transformation takes place through an edge.
Each graph element, either a vertex or an edge, is associated with a  triplet of the form $<Impl,ExecEng,Config>$, either explicitly or implicitly. The $Impl$ property denotes the task or edge implementation, $ExecEng$ provides the engine that will execute each element; and finally, $Config$ captures the configuration of the execution environment, such as the bandwidth reserved for a data transfer across a graph edge, or the number of reducer slots in a Hadoop cluster. Any optimization technique covered in this survey impacts on either the set of $V$ or $E$, or on (part of) the associated triplets.

Data flow optimization is a multi-dimensional problem and its multiple dimensions are broadly divided according to the two flow specification levels. Consequently, we identify the optimization of the \emph{high-level} (or \emph{logical}) flow plan and the \emph{low-level} (or \emph{physical}) flow plan, and each type of optimization mechanism can affect the set of $V$ or $E$ of the workflow graph and their properties.

\begin{figure}[tb!]
\centering
\includegraphics[width=\columnwidth]{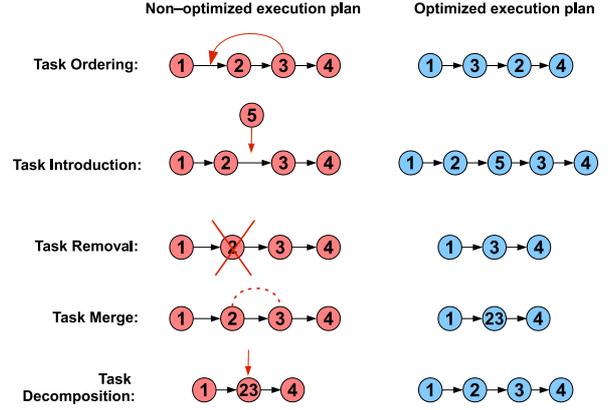}
\caption{Schematic representation of high-level flow optimizations.}
\label{fig:logical}
\end{figure}

The problem of the logical data flow optimization is to define the exact sets $V$ and $E$, so that an objective function is optimized. As such, the logical flow optimization types are largely based on workflow structure reformations, while preserving any dependency constraints between tasks; structure reformations are reflected as modifications in $V$ and $E$.   The output of the optimized flow needs to be semantically equivalent as the output of the initial flow, which practically means that two flows receive the same input data and produce the same output data without considering the way this result was produced. Given that data manipulation takes place only in the context of tasks, logical flow optimization is task-oriented. The logical optimization types are characterized as follows (summarized also in Figure \ref{fig:logical}):

\begin{itemize}
\item \emph{Task Ordering}, where we change the sequence of the tasks by applying a set of partial (re)orderings.  Task (re)ordering affects the set of $E$ of the workflow $DAG$.

\item \emph{Task Introduction}, where new tasks are introduced in the data flow plan in order, for example, to minimize the data to be processed and thus, the overall execution cost. The changes occurred by introducing tasks increase the set of $V$ of the flow graph, which also affects the set $E$, so that the new vertices are connected to the graph.

\item \emph{Task Removal}, which can be deemed as the opposite of task introduction. A task can be safely removed from the flow, if it does not actually contribute to its result dataset. As in the previous case, task removal impacts both on the set $V$, which is reduced, and on $E$ to remove corresponding edges.

\item \emph{Task Merge} is the optimization action of grouping flow tasks into a single task without changing the semantics, applying changes to the set of $V$ in order, for example, to minimize the overall flow execution cost or to mitigate the overhead of enacting multiple tasks.

\item \emph{Task Decomposition}, where a set of grouped tasks is splitted to more than one flow tasks with less complex functionality for generating more optimal sub-tasks. This is the opposite operation of merge action and may provide more optimization opportunities, as discussed in \cite{HPSR12,SWCD12}, because of the potential increase in the number of valid (re)orderings. Similar to the task introduction and merge mechanisms, the optimized workflow plan differs in $V$ with regards to the initial workflow graph, while $E$ is also modified only to reflect changes in $V$.
\end{itemize}

\begin{figure}[tb!]
\centering
\includegraphics[width=\columnwidth]{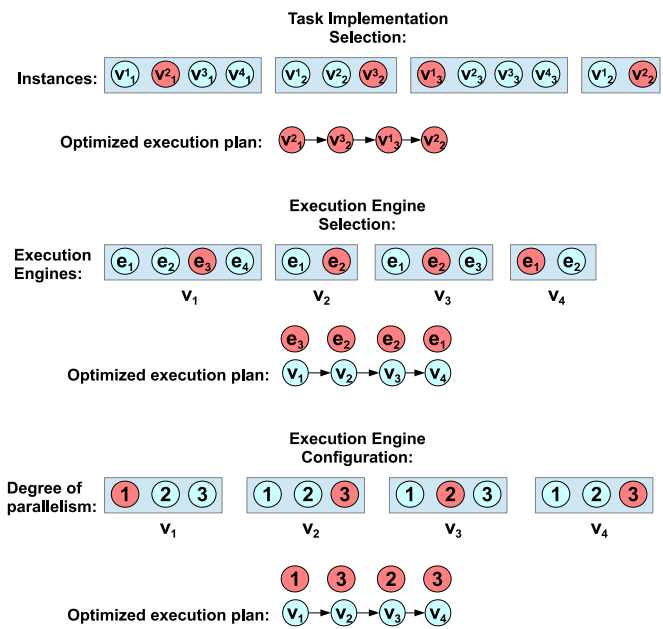}
\caption{Schematic representation of low-level flow optimizations.}
\label{fig:physical}
\end{figure}

At the low level, a wide range of implementation aspects need to be specified so that the flow can be later executed. These aspects are captured by the $<Impl,ExecEng,Config>$ triplet, for each property of which, we identify a different physical data flow optimization type, as follows (see also Figure \ref{fig:physical}):

\begin{itemize}
\item \emph{Task Implementation Selection}, which is one of the most significant lower-level problems in flow optimization. This optimization type includes the selection of the exact, logically equivalent, task implementation for each task that will satisfy the defined optimization objectives~\cite{SWCD12}. A well-known counterpart in database optimization is choosing the exact join algorithm (e.g., hash-join, sort-merge-join, nested loops). In this optimization mechanism case, the $Impl$ property of one or more task or edges have to be specified or modified.
\item \emph{Execution Engine Selection}, where we have to decide the type of processing engine to execute each task. The need for such optimization stems from the availability of multiple options in modern data-intensive flows \cite{KGfgcs14,SWD13}. Common choices, nowadays, include DBMSs, massively parallel engines, such as Hadoop clusters, apart from the execution engines that are bundled with data flow management systems. The corresponding decisions  affect the $ExecEng$ property in the workflow graph.
\item \emph{Execution Engine Configuration}, where we decide on configuration details of the execution environment, such as the bandwidth, CPU, memory to be reserved during execution or the number of cores allocated \cite{SWDH13}. This optimization mechanism refers to the specification of the $Config$ property.
\end{itemize}

\subsection{Optimization Objectives}
An optimization problem can be defined as either \emph{single} or \emph{multiple objective} one depending on the number of criteria that considers. The optimization objectives that are typically presented in the state-of-the-art include the following: \emph{performance}, \emph{reliability}, \emph{availability}, and \emph{monetary cost}. The latter is important when the flow is executed on resources provided at a price, as in public clouds. Other quality metrics can be applied as well (denoted as \emph{other QoS} in \ref{fig:taxonomy}).

The first two objectives require further elaboration. Performance can be defined in several forms, depending, for example, on whether the target is the minimization of the response time, or the resource consumption. The detailed definitions of the performance objective in data flows include the following: minimization of the sum of the task and edge costs \emph{(Sum Cost)}, minimization of the sum of the task and edge costs along the flow critical path \emph{(Critical Path)}, minimization of the most expensive task cost in order to alleviate bottleneck problems \emph{(Bottleneck)}, and maximization of the throughput \emph{(Throughput)}. Each of these definitions may be formally expressed as an objective function, as presented later.

Analogously, reliability may appear in several forms. In our context, reliability reflects how much confidence we have in a data flow execution plan to complete successfully. However, in data flow optimization proposals, we have also encountered the following two reliability aspects playing the role of optimization objectives: \emph{trustworthiness} of a flow (Trust), which is typically based on the trustworthiness of the individual tasks and avoidance of dishonest providers, that is providers with bad reputation; and \emph{Fault Tolerance}, which allows the execution of the flow to proceed even in the case of failures.

\subsection{Optimization Solution Types}
The optimization techniques that have been proposed
constitute \emph{accurate}, \emph{approximate}  or \emph{heuristic} solutions.
Such solutions make sense only when considered in parallel with the complexity of the exact problem they aim to solve. Unfortunately, a big set of the problems in flow optimization are intractable. For such problems, in the case of accurate solutions, a scalable technique cannot be provided.
In the case of approximate optimization solutions, we typically tackle intractable problems in a scalable way while being able to provide guarantees on the approximation bound. Finally, in the last category, we exploit knowledge about the specific problem characteristics and propose algorithms that are fast and exhibit good behavior in test cases, without examining the deviation of the solution from the optimal in a  formal manner.

\subsection{Adaptivity of Data-Centric Flow}
Data flow adaptivity refers to the ability of technique to reoptimize the data flow plan during the execution phase. So, we characterize the optimization techniques as either \emph{static}, where once the flow execution plan is derived it is executed in its entirety, or \emph{dynamic}, where the flow execution plan may be revised on the fly.

\subsection{Execution Environment}
The techniques that are proposed for data flow optimization problem differ significantly according to the execution environment assumed. The execution environment is defined by the type of resources that execute the flow tasks. Specifically, in a \emph{centralized execution environment}, all the tasks of a flow are executed by a single-node execution engine. Additionally, in a \emph{parallel execution environment}, the tasks are executed in parallel by an engine on top of a homogeneous cluster, while in a \emph{distributed execution environment}, the tasks are executed by remote  and potentially heterogeneous execution engines, which are interconnected through ordinary network. Typically, optimizations on the logical level are agnostic to the execution environment, contrary to the physical optimization ones.

\subsection{Metadata}

The set of metadata includes the information needed to apply the optimizations and as such, can be regarded as existential pre-conditions that should hold.
The most basic input requirement of the optimization solutions is an initial set $V$ of tasks. However,
additional metadata with regards to the flow graph are typically required as well. These metadata are both qualitative and quantitative (statistical), as discussed below.
Qualitative metadata include:
\begin{itemize}
\item \emph{Dependencies}, which explicitly refer to the definition of which vertices in the graph should always precede other vertices. Typically, the definition of dependencies comes in the form of an auxiliary graph.

\item \emph{Task schemata}, which refer to the definition of schema of the data input and/or output of each task. Note that dependencies may be produced by task schemata through simple processing \cite{SimVS05}, especially if they contain information about which schema elements are bound or free\cite{KG13}. However, task schemata may serve additional purposes than deriving dependencies, e.g., to check whether a task contributes to the final desired output of the flow.
\item \emph{Task profile}, which refers to information about the execution logic of the task, that is the manner it manipulates its input data; e.g, through analysis of the commands implementing each task. If there is no such metadata, the task is considered as a black-box. Otherwise, information e.g., about which attributes are read and which are written, can be extracted.
\end{itemize}

\noindent
Quantitative metadata include:
\begin{itemize}
\item \emph{Vertex cost}, which typically refers to the time cost, but can also capture other types of costs, such as monetary cost.
\item \emph{Edge cost}, which refers to the cost associated with edges, such as data transmission cost between tasks.
\item \emph{Selectivity}, which is defined as the~(average)~ratio of the output to the input data size of a task and its knowledge is equivalent to estimating the data sizes consumed and produced by each task; sizes are typically measured either in bytes or in number of records (cardinality).
\item \emph{QoS properties}, such as values denoting the task availability, reliability, security, and so on.
\item \emph{Engine details}, which cover issues, such as memory capacity, execution platform configurations, price of cloud machines, and so on.
\end{itemize}

\subsection{Application Domain}
The final dimension across, which we classify existing solutions, is the application domain assumed when each technique is proposed. This dimension sheds light into differentiating aspects of the techniques with regards to the execution environment and the data types processed that cannot be captured by the previous dimensions. Note that the techniques may be applicable to arbitrary data flows in additional application domains than those initially targeted. In this dimension, we consider two aspects:  (i) \emph{domain} of initial proposal, which can be one of the following: ETL flows, data integration, Web Services (WSs) workflows, scientific workflows, MapReduce flows, business processes, database queries or generic; (ii)  \emph{online} (e.g., real-time) vs. \emph{batch} processing. Generic domain proposals aim to a broader coverage of data flow applications, but due to their genericity, they make miss some optimization opportunities that a specific domain proposal could exploit. Also, online applications require more sophisticated solutions, since data is typically streaming and employ additional optimization objectives, such as reliability and acquiring responses under pressing deadlines.


\begin{table*}[tb!]
\tiny
\center
\caption{A summary of the main techniques for producing an optimized flow regarding the dimensions: \emph{mechanisms}, \emph{objectives}, \emph{solution types}, and \emph{metadata}.}
\label{tab:opt1}
\begin{tabular}{|c||c|c|c|c|}
\hline
\textbf{} \textbf{(Refs.,Year)} & \textbf{Mechanisms} & \textbf{Objectives} & \textbf{Solution Types} & \textbf{Metadata} \\ \hline
\mr{(\cite{IBM08},2008),\\(\cite{INFO07},2007)} & \mr{Merge,\\Engine Selection} & Performance & Heuristic & \mr{Task\\Profile} \\ \hline
(\cite{ABDR12},2012) & Ordering & \mr{Performance (Bottleneck/\\Critical Path)} & Accurate ($O(n^6)$) & \mr{Dependencies,\\Vertex Cost,\\Selectivity} \\ \hline
\mr{(\cite{BCDM08},2008)\\ extending \cite{Sriv06}} & \mr{Ordering,\\Implementation Selection} & Performance & Heuristic & \mr{Dependencies,\\Task Schemata, \\Vertex Cost,\\Selectivity} \\ \hline
(\cite{CS99},1999) & Ordering & Performance(Sum Cost) & Approximate & \mr{Vertex Cost,\\Selectivity} \\ \hline
(\cite{ChenZ09},2009) & \mr{Implementation Selection} & \mr{Performance (Critical Path),\\Monetary Cost,\\Reliability} & Heuristic & \mr{Vertex Cost,\\QoS properties} \\ \hline
(\cite{BCGM14},2014) & Removal & Performance & Heuristic ($O(n^2)$) & Task Schemata \\ \hline
(\cite{CGD+15},2015) & \mr{Engine Configuration} & Performance & Heuristic & \mr{Task profile} \\ \hline
(\cite{DSSB05},2005) & Removal & Performance & Heuristic  & \mr{Dependencies,\\Task Schemata} \\ \hline
(\cite{DH12},2012) & Ordering & Performance (Throughput) & Accurate ($O(n^3)$) & \mr{Dependencies,\\Vertex Cost,\\Selectivity} \\ \hline
(\cite{Hel98},1998) & Ordering & Performance(Sum Cost) & Approximate & \mr{Vertex Cost,\\Selectivity} \\ \hline
(\cite{HBTR15},2015) & Engine Configuration & Performance & Heuristic  &  \mr{Task Profile,\\Engine Details} \\ \hline
\mr{(\cite{HJBMY15},2015) \\ extending \cite{HBY13}} & \mr{Task Introduction\\Engine Selection/\\Configuration} & \mr{Performance,\\Monetary Cost,\\Reliability(Fault Tolerance)} & Accurate (exponential) & \mr{Vertex Cost,\\Engine Details} \\ \hline
\mr{(\cite{HPSR12},2012),\\(\cite{RHHLN15},2015)} & \mr{Ordering, \\Introduction/Removal, \\ Decomposition} & Performance (Sum Cost) & \mr{Accurate (exponential)} & \mr{Task Schemata/Profile,\\Vertex Cost,\\Selectivity} \\ \hline
(\cite{KSTI11},2011) & Engine Configuration & \mr{Performance (Sum Cost),\\Monetary Cost} & Heuristic & Vertex Cost \\ \hline
\mr{(\cite{KougkaG15},2015),\\(\cite{KG14},2014)} & Ordering & \mr{Performance (Sum Cost)} & \mr{Accurate (exponential),\\Approximate ($O(n^2)$)} & \mr{Dependencies,\\Vertex Cost, \\Selectivity} \\ \hline
(\cite{KGfgcs14},2014) & \mr{Engine Selection} & Performance (Sum Cost) & Heuristic ($O(n)$) & \mr{Dependencies\\Vertex/Edge Cost} \\ \hline
(\cite{KK10},2010) & Ordering & Performance (Sum Cost) & Approximate ($O(n^2)$) & \mr{Task Schemata,\\Vertex Cost,\\Selectivity} \\ \hline
(\cite{KSP13},2013) & \mr{Implementation Selection, \\ Engine Configuration} & Performance, Other QoS & Heuristic ($O(n)$) & \mr{Vertex Cost,\\ QoS properties} \\ \hline
(\cite{KTMLV08},2008) & \mr{Implementation Selection} & \mr{Performance,\\Availability,\\Monetary Cost} & Heuristic ($O(n)$) & \mr{Vertex Cost,\\ QoS properties} \\ \hline
(\cite{LHB12},2012) & \mr{Merge, \\Engine Configuration} & Performance & Heuristic & \mr{Vertex Cost,\\Task Schemata,\\Selectivity,\\ Engine Details} \\ \hline
(\cite{LI15},2015) & Engine Configuration & Performance & Heuristic & \mr{Vertex Cost,\\Task Profile} \\ \hline
(\cite{SZLCLW14},2014) & Engine Configuration & Performance & Exhaustive & \mr{Vertex Cost,\\Engine Details} \\ \hline
(\cite{SimVS05},2005) & \mr{Ordering, \\ Merge} & \mr{Performance (Sum Cost)} & \mr{Accurate (exponential),\\Heuristic ($O(n^2)$)} & \mr{Vertex Cost,\\Task Schemata} \\ \hline
\mr{(\cite{SWCD12},2012),\\(\cite{SWD13},2013),\\(\cite{SWDH13},2013)}  & \mr{Ordering,\\Decomposition,\\Engine/\\Implementation Selection} & \mr{Performance\\(Constr. Sum Cost\\Bottleneck),\\ Reliability (Fault Tolerance)} & \mr{Accurate (exponential),\\Heuristic ($O(n^2)$)\\} & \mr{Task Schemata,\\Vertex Cost} \\ \hline
\mr{(\cite{SWDC10},2010)\\extending \cite{SimVS05}}& \mr{Ordering,\\Merge,\\Introduction,\\Implementation Selection,\\Engine Configuration} & \mr{Performance\\(Constr. Sum Cost\\Bottleneck),\\Reliability (Fault Tolerance)} & Heuristic ($O(n^2)$) & \mr{Task Schemata,\\Vertex Cost} \\ \hline
(\cite{Sriv06},2006) & Ordering & Performance (Bottleneck) & Accurate ($O(n^5)$) & \mr{Dependencies,\\Vertex Cost,\\Selectivity} \\ \hline
(\cite{TSL+12},2012) & \mr{Implementation Selection} & \mr{Performance,\\Monetary Cost,\\Reliability} &  Heuristic ($O(n)$) & Vertex Cost \\ \hline
(\cite{TGMan11,TGM11}, 2011) & Ordering & Performance (Bottleneck) & Heuristic (exponential) & \mr{Dependencies,\\Vertex/Edge Cost,\\Selectivity} \\ \hline
(\cite{Tziov07},2007)& \mr{Implementation Selection,\\Task Introduction} & Performance (Sum Cost) & \mr{Accurate (exponential)} & \mr{Vertex cost} \\ \hline
(\cite{VSSM07},2007) & Merge & Performance & Heuristic & Task Profile \\ \hline
(\cite{VHA05},2005) & \mr{Implementation Selection} & \mr{Performance,\\Availability,\\Reliability (Trust)} &  Heuristic ($O(n)$) & \mr{Vertex Cost,\\ QoS properties} \\ \hline
(\cite{YLUG99},1999) & Ordering & Performance (Sum Cost) & Approximate ($O(n^2)$) & \mr{Task Schemata,\\Vertex Cost} \\ \hline
(\cite{ZHL15}, 2015) & Engine Selection & \mr{Performance,\\Monetary Cost} & Heuristic & \mr{Vertex Cost,\\Engine details} \\ \hline
\end{tabular}
\end{table*}

\begin{table*}[tb!]
\scriptsize
\center
\caption{A summary of the main techniques for producing an optimized flow regarding the dimensions: \emph{adaptivity}, \emph{execution environment}, and \emph{application domain}.}
\label{tab:opt2}
\begin{tabular}{|c||c|c||c|c|c||c|}
\hline
\textbf{} \textbf{(Refs.,Year)} & \multicolumn{2}{c||}{\textbf{Adaptivity}} & \multicolumn{3}{c||}{\textbf{Execution Environment}} & \textbf{\mr{Application\\Domain}} \\ \hline
 & \textbf{Static} & \textbf{Dynamic} & \textbf{Centralized} & \textbf{Parallel} & \textbf{Distributed} & \\ \hline
\mr{(\cite{IBM08},2008),\\(\cite{INFO07},2007)} & $\bigstar$ & - & $\bigstar$ & - & - & ETL (Batch)\\ \hline
(\cite{ABDR12},2012) & $\bigstar$ & - & - & $\bigstar$ & - & Queries (Online)\\ \hline
\mr{(\cite{BCDM08},2008)\\ extending \cite{Sriv06}} &  $\bigstar$ & - & - & - &  $\bigstar$ & Web Services (Online)\\ \hline
(\cite{CS99},1999) &  $\bigstar$ & - &  $\bigstar$ &-  & - & Queries (Batch)\\  \hline
(\cite{ChenZ09},2009) & $\bigstar$ & - & - & - & $\bigstar$ & Web Services (Batch)\\ \hline
(\cite{BCGM14},2014) & $\bigstar$ & - & $\bigstar$ & - & - & \mr{Scientific\\Workflows (Batch)}\\ \hline
(\cite{CGD+15},2015) & $\bigstar$ & - & - & $\bigstar$ & - & Generic\\ \hline
(\cite{DSSB05},2005) & - & $\bigstar$ & - & $\bigstar$ & - & \mr{Scientific\\Workflows (Batch)}\\ \hline
(\cite{DH12},2012) & $\bigstar$ & - & - & $\bigstar$ & - & \mr{Queries (Online)}\\ \hline
(\cite{Hel98},1998) &  $\bigstar$ & - &  $\bigstar$ &-  & - & Queries (Batch)\\  \hline
(\cite{HBTR15},2015) & - & $\bigstar$ & - & $\bigstar$ & - & Map Reduce (Batch)\\ \hline
\mr{(\cite{HJBMY15},2015)\\ extending \cite{HBY13}} & $\bigstar$ & - & - & - & $\bigstar$ & \mr{Scientific\\Workflows (Batch)}\\ \hline
\mr{(\cite{HPSR12},2012),\\(\cite{RHHLN15},2015)} & $\bigstar$ & - & - & $\bigstar$ & - & \mr{Scientific\\Workflows (Batch)}\\ \hline
(\cite{KSTI11},2011) & $\bigstar$ & - & - & - & $\bigstar$ & Scientific (Online)\\ \hline
\mr{(\cite{KougkaG15},2015),\\(\cite{KG14},2014)} & $\bigstar$ & - & $\bigstar$ & - & - & Generic\\ \hline
(\cite{KGfgcs14},2014) & $\bigstar$ & - & - & - & $\bigstar$  & Generic\\ \hline
(\cite{KK10},2010) & $\bigstar$ & - & $\bigstar$ & - & - & ETL (Batch)\\ \hline
(\cite{KSP13},2013) & - & $\bigstar$ & - & - & $\bigstar$ & Generic\\ \hline
(\cite{KTMLV08},2008) & $\bigstar$ & - & - & - & $\bigstar$ & Web Services (Online)\\ \hline
(\cite{LHB12},2012) & $\bigstar$ & - & - & $\bigstar$ & - & Map Reduce (Batch)\\ \hline
(\cite{LI15},2015) & $\bigstar$ & - & - & $\bigstar$ & - & ETL (Batch)\\ \hline
(\cite{SZLCLW14},2014) & $\bigstar$ & - & - & $\bigstar$ & - & MapReduce (Batch)\\ \hline
(\cite{SimVS05},2005) & $\bigstar$ & - & $\bigstar$ & - & - & ETL (Batch)\\ \hline
\mr{(\cite{SWCD12},2012),\\(\cite{SWD13},2013),\\(\cite{SWDH13},2013)} & $\bigstar$ & - & - & - & $\bigstar$ & ETL (Online)\\ \hline
\mr{(\cite{SWDC10},2010)\\extending \cite{SimVS05}} & $\bigstar$ & - & - & $\bigstar$ & - & ETL (Online)\\ \hline
(\cite{Sriv06},2006) & $\bigstar$ & - & - & - & $\bigstar$ & Web Services (Online)\\ \hline
(\cite{TSL+12},2012) & $\bigstar$ & - & - & - & $\bigstar$ & Generic\\ \hline
(\cite{TGMan11,TGM11}, 2011) & $\bigstar$  & - & - & - & $\bigstar$ & Web Services (Online)\\ \hline
(\cite{Tziov07},2007)& $\bigstar$ & - & $\bigstar$ & - & - & ETL (Batch)\\ \hline
(\cite{VSSM07},2007) & $\bigstar$ & - & - & $\bigstar$ & - & \mr{Business\\Processes (Batch)}\\ \hline
(\cite{VHA05},2005) & $\bigstar$ & - & - & - & $\bigstar$ & Web Services (Online)\\ \hline
(\cite{YLUG99},1999) & $\bigstar$ & - & - & - & $\bigstar$ & \mr{Data\\Integration (Online)}\\ \hline
(\cite{ZHL15}, 2015) & $\bigstar$ & - &  - & - & $\bigstar$ & Generic\\ \hline
\end{tabular}
\end{table*}


\section{Presentation of Existing Solutions}
\label{sec:classification}
Here, we describe the main techniques grouped according to the optimization mechanism. This type of presentation facilitates result synthesis. Grouping by mechanism makes it easier to reason as to whether different techniques employing the same mechanism can be combined or not, e.g., because the make incompatible assumptions. Additionally, the solutions for each mechanism are largely orthogonal to the solutions for another mechanism, which means that, in principle, they can be combined at least in a naive manner. Therefore, our presentation approach provides more insights into how the different solutions can be synthesized.

The discussion is accompanied by a summary of each proposal in Table \ref{tab:opt1} for the dimensions of \emph{mechanisms}, \emph{objectives}, \emph{solution types}, and \emph{metadata}, and Table \ref{tab:opt2}, for the \emph{adaptivity}, \emph{execution environment}, and \emph{application domain} dimensions. When an optimization proposal comes in the form of an algorithm, we also provide the time complexity with respect to the size of the set of vertices $|V|=n$. However, the interpretation of such complexities requires special attention, when there are several other variables of the problem size, as is common in techniques employing optimization mechanisms at the physical level; details are provided within the main text.
The first column of the table mentions also the publication year of each proposal, in order to facilitate the understanding of the proposal's setting and the time evolution of flow optimization.

\begin{figure}[tb!]
\centering
\includegraphics[width=0.8\columnwidth]{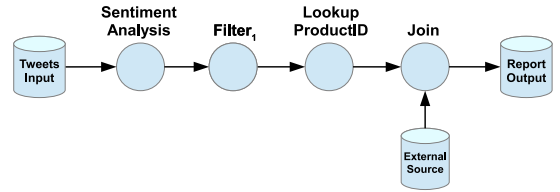}
\caption{A data flow processing Twitter posts.}
\label{fig:analysisExamp}
\end{figure}

Finally, we use a simple running example to present the application of the mechanisms. Specifically, as shown in Figure \ref{fig:analysisExamp}, we consider a data flow that (i) retrieves Twitter posts containing product tags (\emph{Tweets Input}), (ii) performs sentiment analysis (\emph{Sentiment Analysis}), (iii) filters out tweets according to the results of this analysis (\emph{$Filter_1$}), (iv) extracts the product to which the tweet refers to (\emph{Lookup ProductID}), and (v) accesses a static external data source with additional product information (\emph{Join} with \emph{External Source}) in order to produce a report (\emph{Report Output}). In this simple example, in any valid execution plan step (ii) should precede step (iii) and step (iv) should precede step (v).

\subsection{Task Ordering}
\label{sec:ordering}

\begin{table*}
\scriptsize
\center
\caption{A summary of the objective functions in task ordering.}
\label{tab:OF}
\begin{tabular}{|l|l|l|}
\hline
\textbf{Description} & \textbf{Objective Functions} & \textbf{Refs.}\\ \hline
Sum cost & min $\sum c(v_i)$, where $i=1\dots n$ & \cite{HPSR12,KG14,KK10,RHHLN15,SimVS05,YLUG99}\\ \hline
Constrained sum cost & min $\sum c(v_i)$, where $i=1\dots n$ and $g(v_i)<0$ & \cite{SWCD12,SWD13,SWDC10,SWDH13}\\ \hline
\multirow{2}{*}{Bottleneck cost} & min $max(c(v_i))$, where $i=1\dots n$ & \cite{ABDR09,ABDR12,Sriv06}\\ \cline{2-3}
 & min $max(c(v_i,e_{ij}))$, where $i=1\dots n$ & \cite{TGM11,TGMan11} \\ \hline
Critical path cost & min $\sum c(v_i)$, where $v_i$ belongs to \emph{critical path} & \cite{ABDR09,ABDR12} \\ \hline
Throughput & max $\sum f(v_i)$, where $i=1\dots n$ & \cite{DH12}\\ \hline
\end{tabular}
\end{table*}

The goal of \emph{Task Ordering} is typically specified as that of optimizing an objective function, possibly under certain constraints. A common feature of all proposals is that they assign  a metric $m(v_i)$ to each vertex $v_i \in V, i=1\dots n$.
To date, task ordering techniques have been employed to optimize performance. More specifically, all aspects of performance that we introduced previously have been investigated: the minimization of the sum of execution costs of either all tasks (both under and without constraints) or the tasks that belong to the critical path, the minimization of the maximum task cost, and the maximization of the throughput. Table \ref{tab:OF} summarizes the objective functions of these metrics that have been employed by approaches to task ordering in data flow optimization to date. Existing techniques can be modeled at an abstract level uniformly as follows. The metric $m$ refers either to costs (denoted as $c(v_i)$) or to throughput values (denoted as $f(v_i)$). Costs are expressed in either time or abstract units, whereas throughput is expressed as number of records (or tuples) processed per time unit.  A more generic modeling assigns a cost to each vertex $v_i$ along with its outcoming edges $e_{ij},~j=1\dots n$ (denoted as $c(v_i,e_{ij})$).

These objective functions correspond to problems with different algorithmic complexities. Specifically, the problems that target the minimization of the sum of the vertex cost  are intractable \cite{BMS05}. Moreover, Burge et al. \cite{BMS05} discuss that \emph{``it is unlikely that any polynomial time algorithm can approximate the optimal plan to within a factor of $O(n^\theta)$''}, where $\theta$ is some positive constant. The generic bottleneck minimization problem is intractable as well \cite{TGM10}. However, the bottleneck minimization based only on vertex costs and the other two objective functions can be optimally solved in polynomial time \cite{ABDR09,DH12,Sriv06}.

Independently of the exact optimization objectives, all the known optimization techniques in this category assume the existence of dependency constraints between the tasks either explicitly or implicity through the definition of task schemata. For the cost or throughput metadata, some techniques rely on the existence of lower-level information, such as selectivity (see Section \ref{sec:costModel}).

\subsubsection{Techniques for Minimizing the Sum of Costs}
Regarding the minimization of the sum of the vertex costs (first row in Table \ref{tab:OF}), there have been proposed both accurate and heuristic optimization solutions dealing with this intractable problem; apparently the former are not scalable. An accurate task ordering optimization solution is the application of the dynamic programming; dynamic programming is extensively used in query optimization \cite{SACLP79} and such a technique has been proposed for generic data flows in \cite{KG14}. The rationale of this algorithm is to calculate the cost of task subsets of size $n$ based on subsets of size $n-1$. For each of these subsets, we keep only the optimal solution that satisfies the dependency constraints. This solution has exponential complexity even for simple linear non-distributed flows ($O(2^n)$) but, for small values of $n$, is applicable and fast.

Another optimization technique is the exhaustive production of all the topological sortings in a way that each sorting is produced from the previous one with the minimal amount of changes \cite{VR81}; this approach has been also employed to optimize flows in \cite{KougkaG15,KG14}. Despite having a worst case complexity of $O(n!)$, it is more scalable than dynamic programming solution, especially, for flows with many dependency constraints between tasks.

Another exhaustive technique is to define the problem as a state space search one \cite{SimVS05}. In such a space, each possible task ordering is modeled as a distinct state and all states are eventually visited. Similar to the optimization proposals described previously, this technique is not scalable either.
Another form of task-reordering is when a single input/output task is moved before or after a multi-input or a multi-output task \cite{SimVS05,SWDC10}. An example case is when two copies of a proliferate single input/ output task are originally placed in the two inputs of a binary fork operation and after reordering, are moved after the fork. In such a case, the two task copies moved downstream are merged into a single one. As another example, a single input/output task placed after a multi-input task can be moved upstream; e.g., when a filter task placed after a binary fork is moved upstream to both fork input branches (or to just one, based on their predicates). This is similar to traditional query optimization where a selective operation can be moved before an expensive operation like a join.

The branch-and-bound task ordering technique is similar to the dynamic programming one in that it builds a complete flow by appending tasks to smaller sub-flows. To this end, it examines only sub-flows in terms of meeting the dependency constraints and applies a set of recursive calls until generating all the promising data flow plans employing early pruning. Such an optimization technique has been applied in \cite{HPSR12,RHHLN15} for executing parallel scientific workflows efficiently, as part of a new optimization technique for the development of a logical optimizer, which is integrated into the Stratosphere system \cite{Strato14}, the predecessor of Apache Flink. An interesting feature of this approach is that following common practice from database systems it performs static task analysis (i.e., task profiling) in order to yield statistics and fine-grained dependency constraints between tasks going further from the knowledge that can be derived from simply examining the task schemata.

For practical reasons, the four accurate techniques described above are not a good fit for medium and large flows, e.g., with over 15-20 tasks. In these cases, the space of possible solutions is large and needs to be pruned.
Thus, heuristic algorithms have been presented to find near optimal solutions for larger data flows. For example, Simitsis et al. \cite{SimVS05} propose a technique of task ordering by allowing state transitions, which corresponds to orderings that differ in the ordering of only two adjacent tasks. Such transitions are equivalent to a heuristic, which swaps every pair of adjacent tasks, if this change yields lower cost, always preserving the defined dependency constraints, until no further changes can be applied. This  heuristic, initially proposed for ETL flows, can be applied to parallel and distributed execution environments with streaming or batch input data. Interestingly, this technique is combined with another set of heuristics using additional optimization techniques, such as \emph{task merge}. In general, this heuristic is shown to be capable of yielding significant improvements. Its complexity is $O(n^2)$, but there can be no guarantee for how much its solutions can deviate from the optimal one.

There is another family of techniques that minimizing the sum of the tasks by ordering the tasks based on their rank value defined as $\frac{1-sel(v_i)}{c(v_i)}$, where $sel(v_i)$ is the selectivity of $v_i$. The first examples of these techniques were initially proposed for optimizing queries containing UDFs, while dependency constraints between pairs of a join and UDF are considered \cite{CS99,Hel98}. However, they can be applied in data flows by considering flow tasks as UDFs and performing straightforward extensions. For example, an extended version of \cite{CS99}, also discussed in ~\cite{KG14}, builds a flow incrementally in $n$ steps instead of starting from a complete flow and performing changes. In each step, the next task to be appended is the one with the maximum rank value, for which all the prerequisite tasks have been already included. This results in a greedy heuristic of $O(n^2)$ time complexity.

This heuristic has been extended by Kougka et al. \cite{KougkaG15} with techniques that leverage the query optimization algorithm for join ordering by Krishnamurthy et al. \cite{KBZ86} with appropriate post-processing steps in order to yield novel and more efficient task ordering algorithms for data flows. In \cite{KK10}, a similar rationale is followed with the difference that the execution plan is built from the sink to source task. Both proposals build linear plans, i.e., plans in the form of a chain with a single source and a single sink. These proposals for generic or traditional ETL data flows are essentially similar to the \emph{Chain} algorithm proposed by Yerneni et al. \cite{YLUG99} for choosing the order of accessing remote data sources in online data integration scenarios. Interestingly, in \cite{YLUG99}, it is explained that such techniques are $n$-competitive, i.e., they can deviate from the optimal plan up to $n$ times.

\begin{figure}[!]
\centering
\includegraphics[width=0.8\columnwidth]{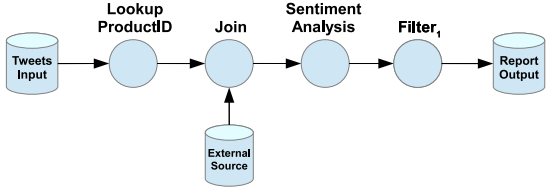}
\caption{An example of optimized task ordering.}
\label{fig:TOexamp}
\end{figure}

The incurred performance improvements can be significant. Consider the example in Figure \ref{fig:analysisExamp}, where let the cost per single input tweet of the five steps be 1, 10, 1, 1, and 5 units, respectively. Let the selectivities be 1, 1, 0.1, 1, and 0.15, respectively. Then the average cost in Figure \ref{fig:analysisExamp} for each initial tweet is 1+10+1+0.1+0.5=12.6, whereas the cost of the flow in Figure \ref{fig:TOexamp} is 1+1+5+1.5+0.15=7.65.
In general, for ordering arbitrary flow tasks in order to minimize the sum of the task costs, any of the above solutions can be used. If the flow is small, exhaustive solutions are applicable; otherwise the techniques in \cite{KougkaG15} are the ones that seem to be capable of yielding the best plans.

Finally, minimizing the sum of the tasks cost appears also in multi-criteria proposals that consider also reliability, in the form of fault tolerance \cite{SWCD12,SWDC10}. These proposals employ a further constraint in the objective function denoted as function $g()$ (see $2^{nd}$ row in Table \ref{tab:OF}). In these proposals, $g()$ defines the number of faults that can be tolerated in a specific time period. The strategy for exploring the search space of different orderings extends the techniques that proposed by  Simitsis et al. \cite{SimVS05}.

\subsubsection{Techniques for Minimizing the Bottleneck Cost} 
Regarding the problem of minimizing the maximum task cost ($3^{rd}$ row in Table \ref{tab:OF}), which acts as the performance bottleneck, there is a \emph{Task Ordering} mechanism initially proposed for the parallel execution of online WSs represented as queries \cite{Sriv06}. The rationale of this technique is to push the selective flow tasks (i.e., those with $sel<1$) in an earlier stage of the execution plan in order to prune the input dataset of each service. Based on the selectivity values, there may be cases where the output of a service may be dispatched to multiple other services for executing in parallel or in a sequence having time complexity in $O(n^5)$ in the worst case. The problem is formulated in a way that it is tractable and the solutions is accurate.

Another optimization technique that considers task ordering mechanism for online queries over Web Services appears in \cite{TGM11,TGMan11}. The formulation in these proposals extends the one proposed by Srivastava et al. \cite{Sriv06} in that it considers also edge costs. This modification renders the problem intractable \cite{TGM10}. The practical value is that edge costs naturally capture the data transmission between tasks in  a distributed setting. The solution proposed by Tsamoura et al. \cite{TGM11,TGMan11} consists of a branch-and-bound optimization approach with advanced heuristics for early pruning and despite of its exponential complexity, it is shown that it can apply to flows with hundreds of tasks, for reasonable probability distributions of vertex and edge costs.

The techniques for minimizing the bottleneck cost can be combined with those for the minimization of the sum of the costs. More specifically, the pipelined tasks can be grouped together and for the corresponding sub-flow, the optimization can be performed according to the bottleneck cost metric. Then, these groups of tasks can be optimized considering the sum of their costs. This essentially leads to a hybrid objective function that aims to minimize the sum of the costs for segments of pipelining operators, where each segment cost is defined according to the bottleneck cost. A heuristic combining the two metrics has appeared in  \cite{SWDC10}.

\subsubsection{Techniques for Optimizing the Critical Path}
A technique that considers the critical path providing an accurate solution  has appeared in \cite{ABDR12}. This work has $O(n^6)$ time complexity and has been initially proposed for online queries in parallel execution environments, but is also applicable to data flows. The strong point of this solution is that it can perform bi-objective optimization combining the bottleneck and the critical path criteria.

\subsubsection{Techniques for Maximizing the Throughput}
Reordering the filter operators of a workflow can be used to find an optimal query execution plan that maximizes throughput leveraging pipelined parallelism. Such a technique has been presented by Deshpande et al. \cite{DH12} considering queries with tree-shaped constraints for parallel execution environment providing an accurate solution that has $O(n^3)$ time complexity. In this proposal, each task is assumed to be executed on a distinct node, where each node has a certain throughput capacity that should not be exceeded. The unique feature of this proposal is that it produces a set of plans that need to be executed concurrently in order to attain throughput maximization. The drawback is that it cannot handle arbitrary constraint graphs, which implies that its applicability to generic data flows is limited.

\subsubsection{Task Cost Models}
Orthogonally to the objective functions in Table~\ref{tab:OF}, different cost models can be employed to derive $c(v_i)$, the cost of the $i^{th}$ task $v_i$. The important issue is that a task cost model can be used as a component in any cost-based optimization technique, regardless of whether it has been employed in the original work proposing that technique.
A common assumption is that $c(v_i)$ depends on the volume of data processed by $v_i$, but this feature can be expressed in several ways:

\label{sec:costModel}
\begin{itemize}
\item $c(v_i) = \prod_{j=1}^{|T_i^{prec}|} sel_j * cpi_i$ : this cost model defines the cost of the  $i^{th}$ task as the product of i) the cost per input data unit  ($cpi_i$) and ii) the product of the selectivities $sel$ of preceding tasks; $T_i^{prec}$ is the set of all the tasks between the data sources and $v_i$. This cost model is explicitly used in proposals such as \cite{KougkaG15,KG13,KG14,KK10,YLUG99}.
\item $c(v_i)=rs(v_i)$ : In this case, the cost model is defined as the size of the results ($rs$) of $v_i$; it is used in \cite{YLUG99}, where each task is a remote database query.
\item $c(v_i) = \alpha_i\cdot CPU(v_i) + \beta_i \cdot IO(v_i) + \gamma_i \cdot Ship(v_i)$: this cost model is a weighted sum of the three main cost components, namely the cpu, I/O, and data shipping costs. Further, $CPU(v_i)$ can be elaborated and specified as $\prod_{j=1}^{|T_i^{prec}|} sel_j * cpi_i$ (defined above) plus a startup cost. I/O costs depends on the cost per input data unit to access secondary storage. Data communication cost $Ship(v_i)$ depends on the size of the input of $v_i$, which, as explained earlier, depends also on previous tasks and the vertex selectivity $sel_i$. $\alpha$, $\beta$, and $\gamma$ are the weights. Such an elaborate cost model has been employed by Hueske et al. \cite{HPSR12}.

\item $c(v_i) = proc(v_i) + part(v_i)$: This cost model is suggested by Simitsis et al. \cite{SWDC10}. It explicitly covers task parallelization and splits the cost of a tasks into the processing cost $proc$ and the cost to partition and merge data $part$. The former cost is divided into a part that depends on input size and a fixed one. The proposal in \cite{SWDC10} treats differently the tasks in the flow that add recovery points or create replicas by providing specific formulas for them.
\end{itemize}

\subsubsection{Additional Remarks}
Regarding the execution environment, since the task (re-)ordering techniques refer to the logical WEP level, they can be applied to both centralized and distributed flow execution environments. However, in parallel and distributed environments, the data communication cost needs to be considered. The difference between these environments with regards to the communication cost is that in the latter, this cost depends both on the sender and receiver task and as such, it needs to be represented, not as a component of vertex cost but as a property of edge cost.

Additionally, very few techniques, e.g. \cite{SimVS05}, explicitly consider reorderings between single input/output and multiple-input or multiple-output tasks; however, this type of optimization requires further investigation in the context of complex flow optimization.

Finally, none of the proposed techniques for task ordering technique discussed are adaptive ones, that is they do not consider workflow re-optimization during its execution phase. In general, adaptive flow optimization is a subarea in its infancy.  However, B{\"{o}}hm et al. \cite{BoehmHL14} has proposed solutions for choosing when to trigger re-optimization, which, in principle, can be coupled with any cost-based flow optimization technique.

\subsection{Task Introduction}

Task introduction has been proposed for three reasons.

\begin{figure}[!]
\centering
\includegraphics[width=0.8\columnwidth]{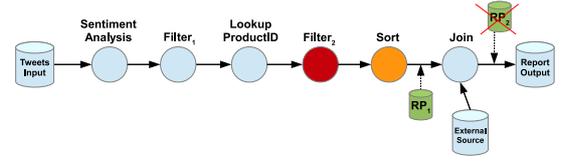}
\caption{Examples of \emph{Task Introduction} techniques.}
\label{fig:TIexamp1}
\end{figure}

Firstly, to achieve fault-tolerance through the introduction of recovery points and replicator tasks in online ETLs \cite{SWDC10}. For recovery points, a new node storing the current flow state is inserted in the flow in order to assist recovering from failures without needing to recompute the flow from scratch. Adding a recovery (to a specific point in the plan) depends on a cost function that compares the projected recovery cost in case of failure against the cost to maintain a recovery point. Additionally, the replicator nodes produce copies of specified sub-flows in order to tolerate local failures, when no recovery points can be inserted, e.g., because the associated overhead increases the execution time above a threshold. In both cases of task introduction, the semantics of the flow are immutable. The proposed technique extends the state space search in \cite{SimVS05} after having pruned the state search space. The objective function employed is the constrained sum cost one ($2^{nd}$ row in Table \ref{tab:OF}), where the constraint is on the number of places where a failure can occur. The cost model explicitly covers the recovery maintenance overhead (last case in Sec. \ref{sec:costModel}). The key idea behind the pruning of search space is first to apply task reordering and then, to detect all the promising places to add the recovery points based on heuristic rules. An example of the technique is in Figure \ref{fig:TIexamp1} and suppose that we examine the introduction of up to two recovery points. The two possible places are just after the  \emph{Sort} and \emph{Join} tasks, respectively. Assume that the most beneficial place is the first one, denoted as $RP_1$. Also, given  $RP_1$, $RP_2$ is discarded because it incurs higher cost than re-executing the \emph{Join} task.
Similarly to the recovery points above, the technique proposed by Huang et al. \cite{HJBMY15} introduces operations that copy intermediate data from transient nodes to primary ones, using a cluster of machines containing both transient and primary cloud machines; the former can be reclaimed by the cloud provided at any time, whereas the latter are allocated to flow execution throughout its execution.

Secondly, task introduction has been employed by Rheinl{\"{a}}nder et al. \cite{RHHLN15} to automatically insert explicit filtering tasks, when the user has not initially introduced them. This becomes plausible with a sophisticated task profiling mechanism employed in that proposal, which allows the system to detect that some data are not actually needed. The goal is to optimize a sum cost objective function, but the technique is orthogonal to any objective function aiming at performance improvement. For example, in Figure \ref{fig:TIexamp1}, we introduce a filtering task if the final report needs only a subset of the initial data, e.g., it refers to a specific range of products.

Third, task introduction can be combined with \emph{Implementation Selection} (Section  \ref{subsec:implementation}). An example appears in \cite{Tziov07}, where the purpose is to exploit the benefit of processing sorted records. To this end, it explores the possibility of introducing new vertices, called sorters,  and then to choose task implementations that assume sorted input; the overhead of the insertion of the new tasks is outweighed by the benefits of sort-based implementations. In Figure \ref{fig:TIexamp1}, we add such a sorter task just before the \emph{Join} if a sort-based join implementation and report output is preferred. Proactively ordering data to reduce the overall cost has been used in traditional database query optimization \cite{dbsystems} and it seems to be profitable for ETL flows as well.

Finally, all these three techniques can be combined; e.g., in the example all can apply simultaneously yielding the complete plan in the figure.

\subsection{Task Removal}
A set of  optimization proposals support the idea of removing a task or a set of tasks from the workflow execution plan without changing the semantics in order to improve the performance; these proposals have been proposed mostly for offline scientific workflows, where it is common to reuse tasks or sub-flows from previous workflows without necessarily examining whether all tasks included are actually necessary or whether some results are already present. Three techniques adopt this rationale \cite{BCGM14,DSSB05,RHHLN15}, which are discussed in turn.

\begin{figure}[!]
\centering
\includegraphics[width=0.8\columnwidth]{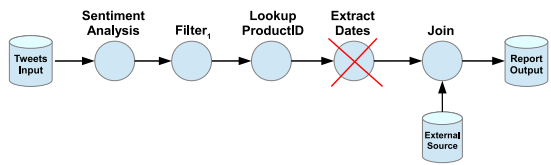}
\caption{An example of the \emph{Task Removal} technique.}
\label{fig:TRexamp}
\end{figure}

The idea of Rheinl{\"{a}}nder et al. \cite{RHHLN15} is to remove a task or multiple tasks until the workflow consists only of tasks that are necessary for the production of the desired output. This implies that the execution result dataset remains the same regardless of the changes that have been applied. It aims to protect users that have carelessly copied data flow tasks from previous flows. In Figure \ref{fig:TRexamp}, we see that, initially, the example data flow contains an \emph{Extract Dates} task, which is not actually necessary.

The heuristic of Deelman et al. \cite{DSSB05} has been proposed for a parallel execution environment and is one of the few dynamic techniques allowing the reoptimization of the workflow during the workflow execution.  At runtime, it checks whether any intermediate results already exist at some node, thus making part of the flow obsolete. Both \cite{RHHLN15} and  \cite{DSSB05} are rule-based and do not target an objective function directly.

Another approach for applying task removal optimization mechanism is to detect the duplicate tasks, i.e., tasks performing exactly the same operation and keep only a single copy in the execution workflow plan \cite{BCGM14}. This might be caused by carelessly combining existing smaller flows from a repository, e.g., myExperiment\footnote{\url{www.myexperiment.org/} in bio-informatics.} A necessary condition in order to ensure that there will be no precedence violations is that these tasks must be dependency constraint free, which is checked with the help of the task schemata. Such a  heuristic  has $O(n^2)$ time complexity.

\subsection{Task Merge}
\emph{Task Merge} has been also employed for improving the performance of the workflow execution plan. The main technique is to apply re-writing rules to merge tasks with similar functions into one bigger task. There are three techniques in this group, all tailored to a specific setting. As such, it is unclear whether they can be combined.

First, in \cite{VSSM07}, tasks that encapsulate invocations to an underlying database are merged so that fewer (and more complex) invocations take place. This rule-based heuristic has been proposed for business processes, for which it is common to access various data stores, and such invocations incur a large time overhead.

Second, a related technique has been proposed for SQL statements in commercial data integration products \cite{IBM08,INFO07}. The rationale of this idea is to group the SQL statements into a bigger query in order to push the task functionalities to the best processing engine. Both approaches presented in \cite{IBM08,INFO07} derive the necessary information about the functionality of each task with the help of task profiling and produce larger queries employing standard database technology. For example, instead of processing a series of SQL queries to transform data, it is preferable to create a single bigger query. As previously, the optimization is in the form of a heuristic that does not target to optimize any objective function explicitly. A generalization of this idea to languages beyond SQL is presented by Simitsis et al. \cite{SWCD12,SWDH13} and a programming language translator has been described by Jovanovic et al. \cite{JSW14a,JSW14}.

Third, Harold et al. \cite{LHB12} presents a heuristic non-exhaustive solution for merging MapReduce jobs. Merging occurs at two levels: first MapReduce jobs are tried to be transformed into Map-only jobs. Then, sharing common Map or Reduce tasks is investigated. These two aspects are examined with the help of a 2-phase heuristic technique.

Finally, in the optimizations in \cite{SimVS05,SWDC10}, which rely on a state space search as described previously, adjacent tasks that should not be separated may be grouped together during optimization. The aim of this type of merger is not to produce a flow execution plan with fewer and more complex tasks (i.e., no actual task merge optimization takes place), but to reduce the search space so that the optimization is speeded-up; after optimization, the merged tasks are split.

\subsection{Task Decomposition}
An advanced optimization functionality is \emph{Task Decomposition}, according to which, the operations of a task are split into more tasks, this results in a modification of the set $V$ of vertices.  This mechanism has appeared in \cite{HPSR12,RHHLN15} as a pre-processing step, before the task ordering takes place. Its advantage is that it opens-up opportunities for ordering, i.e., it does not optimize an objective function in its own but it enables more profitable task orderings.

Task decomposition is also employed by Simitsis et al. \cite{SWCD12,SWD13,SWDH13}. In these proposals, complex analysis tasks, such as sentiment analysis presented in previous examples, can be split into a sequence of tasks at a finer granularity, such as tokenization, and part-of-speech tagging.

Note that both these techniques are tightly coupled to the task implementation platform assumed.

\subsection{Task Implementation Selection}
\label{subsec:implementation}

\begin{figure}[!]
\centering
\includegraphics[width=0.8\columnwidth]{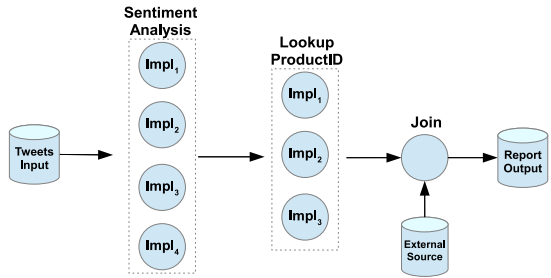}
\caption{An example where \emph{Task Implementation Selection} is applicable, where there are four equivalent ways to implement sentiment analysis and three ways to extract product ids.}
\label{fig:TISexamp1}
\end{figure}

A set of optimization techniques target the \emph{Implementation Selection} mechanism. At a high level, the problem is that there exist multiple equivalent candidate implementations for each task and we need to decide which ones to employ in the execution plan. For example, a task encapsulating a call to a remote WS, can contact multiple equivalent WSs, or a task may be implemented to run both in a single-machine mode or in as a MapReduce program.
These techniques typically require as input metadata the vertex costs of each task implementation alternative. Suppose that, for each task, there are $m$ alternatives. This leads to a total of $O(m^n)$ of combinations; thus a key challenge is to cope with the exponential search space. In general, the number of alternatives for each task may be different and the total number of combinations is the product of these numbers. For example, in Figure \ref{fig:TISexamp1}, there are four and three alternatives ($Impl_1, ..., Impl_n$) for the \emph{Sentiment Analysis}  and \emph{Lookup Product} tasks, respectively, corresponding to twelve combinations.

It is important to note that, conceptually, the choice of the implementation of each task is orthogonal to decisions on task ordering and the rest of the high-level optimization mechanisms. As such, the techniques in this section can be combined with techniques from the previous sections.

A brute force, and thus of exponential complexity approach to finding the optimal physical implementation of each flow task before its execution has appeared in \cite{Tziov07}. This approach models the problem as a state space search one and, although it assumes that the sum cost objective function is to be optimized, it can support other objective functions too. An interesting feature of this solution is that it explicitly explores the potential benefit from processing sorted data.  Also, the ordering and task introduction algorithm in \cite{SWDC10} allows for choosing parallel flavors of tasks. The parallel flavors, apart from cloning the tasks as many times as the degree of partitioned parallelism decided, explicitly consider issues, such as splitting the input data, distributing them across all clones, and merging all their outputs. These issues are reflected in an elaborate cost function as mentioned previously, which is used to decide whether parallelization is beneficial.

Additionally to the optimization techniques above, there is a set of multi-objective optimization approaches for \emph{Implementation Selection}. These multi-objective heuristics, apart from the vertex cost, require further metadata that depend on the specified optimization objectives. For example, several multi-objective optimization approaches have been proposed for flows, where each task is essentially an invocation to an online WS that may not be always available; in such settings, the aim of the optimizer is the selection of the best service  for each service type taking into account both performance and availability metadata.

Three proposals that target this specific environment are \cite{KTMLV08,TSL+12,VHA05}. To achieve scalability, each task is checked  in isolation, thus resulting in $O(nm)$ time complexity, but at the expense of finding local optimal solutions only. Kyriazis et al. \cite{KTMLV08} consider availability, performance, and cost for each task. As initial metadata, scalar values for each objective and for candidate services are assumed to be in place.
The main focus of the proposed solution is (i) on normalizing and scaling the initial values for each of the objectives and (ii) on devising an iterative improvement algorithm for making the final decisions for each task. The multi-objective function is either the optimization of a single criterion under constraints on the others or the optimization of all the objectives at the same time. However, in both cases, no optimality guarantees (e.g., finding a Pareto optimal solution) are provided.

The proposal in \cite{VHA05} is similar in not guaranteeing pareto optimal solutions. It considers performance, availability, and reliability for each candidate WS, where each criterion is weighted and contributes to a single scalar value, according to which services are ordered. The notion of reliability in this proposal is based on its trustworthiness. \cite{TSL+12} is another service selection proposal that considers the three objectives, namely performance, monetary cost, and reliability in terms of successful execution. The service metadata are normalized and the technique proposed employs a max-min heuristic that aims to select a service based on its smallest normalized value. An additional common feature of the proposals in \cite{KTMLV08,TSL+12,VHA05} is that no objective function is explicitly targeted.

Another multi-objective optimization approach to choosing the best implementation selection of each task consists of linear complexity heuristics \cite{KSP13}. The main value of those heuristics are that they are designed to be applied on the fly, thus forming one of the few existing adaptive data flow optimization proposals. Additionally, the technique proposed by Braga et al. \cite{BCDM08} extends the task ordering approach in \cite{Sriv06} so that, for each task, the most appropriate implementation is first selected. None of these proposals employ a specific objective function as well.
Finally,  multi-objective WS selection mechanism can be performed with the help of  ant colony optimization algorithms; an example of applying this optimization technique for selecting WS instantiations between multiple candidates in a setting where the workflows mainly consist of a series of remote WS invocations appears in \cite{ChenZ09}, which is further extended by Tao et al. \cite{Tao14}.

Based on the above descriptions, two main observations can be drawn regarding the majority of the techniques. Firstly, they address a multi-objective problem. Secondly, they are proposed for a WS application domain. The latter may imply that transferring the results to dataflows where tasks exchange big volumes of data directly may not be straightforward.

\subsection{Execution Engine Selection}
The techniques in this category focus on choosing the best execution engine for executing the data flow tasks in distributed environments, where there are multiple options. For example, assume that the sentiment analysis in our running example can take place on either a DBMS server or a MapReduce cluster. As previously, for the techniques using this mechanism, the vertex cost of each task for each candidate execution engine is a necessary piece of metadata for the optimization algorithm. Also, corresponding techniques are orthogonal to optimizations referring to the high-level execution plan aspects.

For those tasks that can be executed by multiple engines, an exhaustive solution can be adopted for optimally allocating the tasks of a flow to different execution engines in order to meet multiple objectives. The drawback is that an exhaustive solution in general does not scale for large number of flow tasks and execution engines similarly to the case of task implementation selection. To overcome this, a set of heuristics can be used for pruning the search space \cite{SWCD12,SWD13,SWDH13}. This technique aims to improve not only the performance, but also the reliability of ETL workflows in terms of fault tolerance. Additionally, a multi-objective solution for optimizing the monetary cost and the performance is to check all the possible execution plans that satisfy a specific time constraint; this approach cannot scale for execution plans with high number of operators. The objective functions are those mentioned in Section \ref{sec:ordering}. The same approach to deciding the execution engine, can be used to choose the task implementation in \cite{SWCD12,SWD13,SWDH13}.

Anytime single-objective heuristics for choosing between multiple engine have been proposed Kougka et al. \cite{KGfgcs14}. Such heuristics take into account, apart from vertex costs, the edge costs and constraints on the capability of an engine to execute certain tasks and are coupled with a dynamic programming pseudo-polynomial algorithm that can find optimal allocation for a specific form of DAG shapes, namely linear ones. The objective function is minimizing the sum of the costs for both tasks and edges, extending the definition in Table \ref{tab:OF}:  min $\sum c(v_i,e_{ij})$, where $i,j=1\dots n$.

A different approach to engine selection has appeared in the commercial tools in  \cite{INFO07,IBM08}. There, the main option is ETL operators to execute on a specialized data integration server, unless a heuristic decides to delegate the execution of some of the tasks to the underlying databases, after merging  the tasks and reformulating them as a single query.

Finally, the engine selection mechanism can be employed in combination with configuration of execution engine parameters. An example technique is presented by Huang et al. \cite{HBY13}, where the initial optimization step deals with the decision of the best type of execution engine and then, the configuration parameters are defined, as it is analyzed in Section \ref{sec:config}. This technique is extended by Huang et al. \cite{HJBMY15}, which focuses on how to decide on the usage of a specific type of cloud machines, namely spot instances. The problem of deciding whether to employ spot instances in clouds is also considered by Zhou et al. \cite{ZHL15}.

\subsection{Execution Engine Configuration}
\label{sec:config}
This type of flow optimization has recently received attention due to the increasing number of parallel data flow platforms, such as Hadoop and Spark. The \emph{Engine Configuration} mechanism can serve as a complementary component of an optimization technique that applies implementation or engine selection, and in general, can be combined with the other optimization mechanisms. For example, the rationale of the heuristic presented by Kumbhare et al. \cite{KSP13} (based on variable sized bin packing) is also to decide the best implementation for each task and then, dynamically configure the resources, such as the number of CPU cores allocated, for executing the tasks. A common feature of all the solutions in this section is that they deal with parallelism, but from different perspectives depending on the exact execution environment.

 A  specific type of engine configuration, namely to decide the degree of parallelism in MapReduce-like clusters for each task and parameters, such as the number of slots on each node, appears in \cite{HBY13}. The time complexity of this optimization technique is exponential. This is repeated for each different type of machines (i.e., different type of execution engine), assuming a context where several heterogeneous clusters are at user's disposal.
Both of these techniques have been proposed for cloud environments and aim to optimize multiple optimization criteria.

In general, execution engines come with a large number of configuration parameters and fine tuning them is a challenging task. For example, MapReduce systems may have more than one hundred configuration parameters. The proposal in \cite{SZLCLW14} aims to provide a principle approach to their configuration. Given the number of MapReduce slots and hardware details, the proposed algorithm initially checks all combinations of four key parameters, such as the number of map and reduce waves, and whether to use compression or not. Then, the values of a dozen other configuration parameters that have significant impact on performance are derived. The overall goal is to reduce the execution time taking to account the pipeline nature of MapReduce execution.

An alternative configuration technique is employed by Lim et al. \cite{LHB12}, which leverages the what-if engine initially proposed by Herodotou et al. \cite{HB11}. This engine is responsible to configure execution settings, such as memory allocation and number of map and reduce tasks, by answering questions on real and hypothetical input parameters using a random search algorithm. What-if analysis is also employed by \cite{HBTR15} for optimally configuring memory configurations. The distinctive feature of this proposal is that it is dynamic in the sense that it can take decisions at runtime leading to task migrations.

In a more traditional ETL setting, apart from the optimizations described previously, an additional optimization mechanism has been proposed by Simitsis et al. \cite{SWDC10} in order to define the degree of parallelism. Specifically, due to the large size of data that a workflow has to process, data is partitioned to be executed following the intra-operator parallelism paradigm. The parallelism is considered profitable whenever the overhead of data partitioning and merging does not incur an overhead higher then the expected benefits. Sometimes it might be worth investigating whether splitting an input dataset into partitions could reduce the latency in ETL flow execution on a single server as well. An example study can be found in \cite{LI15}.



Another approach to choosing the degree of parallelism appears in \cite{KSTI11}, where a set of greedy and simulated annealing heuristics that decide the degree of parallelism are proposed. This proposal considers two objectives, performance and monetary cost assuming that resources are offered by a public cloud at a certain price. The objective function targets either the  minimization of the sum of the task costs constrained by a defined monetary budget, or the minimization of the monetary cost under a constraint on runtime. Additionally, both metrics can be minimized simultaneously using an appropriate objective function, which expresses the speedup when budget is increased.

Another optimization technique in \cite{CGD+15} proposes a set of optimizations at the chip processor level and more specifically, proposes heuristics to drive compiler decisions on whether to execute low-level commands in a pipelined fashion or to employ SIMD (single instruction multiple data) parallelism. Interestingly, these optimizations are coupled with traditional database-like ones at a higher level, such as pushing selections as early as possible.


\section{Evaluation Approaches}
\label{sec:eval}

\begin{table}
\tiny
\center
\caption{Experimental Evaluation of Proposals}
\label{tab:evaluation}
\begin{tabular}{|c|c|c|c|c|}
\hline
\multicolumn{5}{|c|}{{\bf Evaluation}}                                                                                                                                                                           \\ \hline
\textbf{(Refs.,Year)} & \bf{\mr{Workflow\\ Type}} & \bf{\mr{Data\\Type}} & \bf{\mr{Implemen-\\tation\\Envir.}} & \bf{\mr{Max.\\DAG\\Size}}\\ \hline
(\cite{Hel98},1998) & Synthetic & Synthetic & Real & 4 \\ \hline
(\cite{CS99},1999) & Synthetic & Synthetic & Real & 16 \\ \hline
(\cite{YLUG99},1999) & Synthetic  & Synthetic & Simul. & 15\\ \hline
(\cite{SimVS05},2005) & Synthetic & Synthetic & Simul. & 70\\ \hline
(\cite{DSSB05},2005) & Real & Real & Real & N/A\\ \hline
(\cite{VHA05},2005) & Synthetic & Synthetic & Simul. & 200\\ \hline
(\cite{Sriv06},2006) & Synthetic & Synthetic & Real & 4\\ \hline
(\cite{Tziov07},2007) & Synthetic & Synthetic & Simul. & 15\\ \hline
(\cite{VSSM07},2007) & Synthetic & Synthetic & Real & N/A\\ \hline
(\cite{BCDM08},2008) & Real & Real & Simul. & 7\\ \hline
(\cite{KTMLV08},2008) & Synthetic & Synthetic & Real & 8 \\ \hline
(\cite{ChenZ09},2009) & Synthetic & Synthetic & Simul. & 120 \\ \hline
(\cite{KK10},2010) & Synthetic & Synthetic & Simul. & 60\\ \hline
(\cite{SWDC10},2010) & Real & Synthetic & Real  & 80\\ \hline
(\cite{KSTI11},2011) & Real & Synthetic & Real & 500 \\ \hline
(\cite{TGMan11,TGM11},2011) & Synthetic & Synthetic & Simul. & 100\\ \hline
\mr{(\cite{HPSR12},2012),\\(\cite{RHHLN15},2015)} & Real & Real & Real & 15\\ \hline
(\cite{LHB12},2012) & Real & Synthetic & Real & 14 \\ \hline
\mr{(\cite{SWCD12},2012),\\(\cite{SWD13,SWDH13},2013)} & Real & Real & Real & 15\\ \hline
\mr{(\cite{HBY13},2013),\\(\cite{HJBMY15},2015)} & Real & Synthetic & Real & N/A \\ \hline
(\cite{KSP13},2013) & Synthetic & Synthetic & Simul. & 4 \\ \hline
(\cite{KGfgcs14},2014) & Real & Synthetic & Simul. & 200\\ \hline
(\cite{SZLCLW14},2014) & Real & Synthetic  & Real & $<10$  \\ \hline
(\cite{BCGM14},2014) & Real & Real & Real & N/A \\ \hline
(\cite{CGD+15},2015) & Real & Synthetic & Real & N/A \\ \hline
(\cite{HBTR15},2015) & Real & Real & Real & N/A  \\ \hline
\mr{(\cite{KougkaG15},2015),\\(\cite{KG14},2014)} & Synthetic & Synthetic & Simul. & 200   \\ \hline
(\cite{LI15},2015) & Real & Synthetic & Real & 11\\ \hline
(\cite{ZHL15},2015) & Real & Synthetic & Both & $>10000$ \\ \hline
\end{tabular}
\end{table}

Here, we describe the evaluation methods used in each proposed work. We can divide the proposals in three categories.

The first category includes the optimization proposals that are theoretical in their nature and their results are not accompanied by experiments. Examples of this category are \cite{ABDR12,DH12}. The second category consists of optimizations that have found their way into data flow tools; the only examples in this category are~\cite{IBM08,INFO07}.

The third category covers the majority of the proposals, for which experimental evaluation has been provided. We are mostly interested in three aspects of such experiments, namely the \emph{workflow type} used in the experiments, the \emph{data type} used to instantiate the workflows, and the \emph{implementation environment} of the experiments. In Table \ref{tab:evaluation}, the experimental evaluation approaches are summarized, along with the maximum DAG  size (in terms of number of tasks) employed. Specifically, the implementation environment defines the execution environment of a workflow during the evaluation procedure. The environment can be a \emph{real-world} one, which considers either the customization of an existing system to support the proposed optimization solutions or the design of a prototype system, which is essentially a new platform, possibly designed from scratch and tailored to support the evaluation. A common approach consists of a \emph{simulation} of a real execution environment. Discussing the pros and cons of each approach is out of our scope, but in general, simulations allow the experimentation with a broader range of flow types, whereas real experiments can better reveal the actual benefits of optimizations in practice.

As shown in Table \ref{tab:evaluation}, the majority of the optimization techniques have been evaluated by executing workflows in a simulated environment.
The real environments that have been employed are as follows. The techniques in \cite{SWCD12,SWD13,SWDC10,SWDH13} that focused on (complex) ETL data flows have been evaluated with the help of extensions to the Pentaho Data Integration (Kettle) tool, a commercial database, and a MapReduce engine.  The proposals in \cite{HPSR12,RHHLN15} have been tested in the Stratosphere a Big Data Analytics platform \cite{Strato14}. A MapReduce-inspired prototype, called Cumulon, is used for the evaluation of the techniques in \cite{HBY13,HJBMY15}. Other MapReduce extensions have been employed in \cite{HBTR15,LHB12,SZLCLW14}. To evaluate techniques initially proposed for flows consisting of calls to WSs, both ad-hoc prototypes \cite{KTMLV08,Sriv06} and extensions to engines, such as Taverna \cite{BCGM14} and Web-Sphere Process Server \cite{VSSM07} have been used. Part of the evaluation of \cite{ZHL15} involved running Pegasus on a public cloud. The techniques in \cite{CGD+15} and \cite{KSTI11} are part of broader prototype systems, called Tupleware and ADP, respectively. Finally, the early works on database queries including UDFs were implemented in a DBMS \cite{CS99,Hel98}.

The type of the workflows considered are either synthetic or real-world. In the former case, arbitrary DAGs are produced, e.g., based on the guidelines in \cite{VKTS07}. In the latter case, the flow structure is according to real-world cases. For example, the evaluation of \cite{ChenZ09,BCGM14,DSSB05,KSTI11,KGfgcs14,ZHL15} is based on real-world scientific workflows, such as the Montage and Cybershake ones described in \cite{Juve13}. Another example of real-world workflows are derived by  TPC-H queries (used for some of the evaluation experiments in \cite{HPSR12,LHB12,RHHLN15} along with real world text mining and information extraction examples). In \cite{SWCD12,SWD13,SWDC10,SWDH13}, the evaluation of the optimization proposals is based on workflows that represent arbitrary, real-world data transformations and text analytics. The case studies in \cite{CGD+15,LHB12} include standard analytical algorithms, such as PageRank, k-means, logistic regression, and naive bayes.

The datasets used for workflow execution may affect the evaluation results, since they specify the range of the statistical metadata considered. The processed datasets can be either synthetic or real ones extracted by repositories, such as the Twitter repository with sample data of real tweets.  Examples of real datasets used in \cite{HPSR12,RHHLN15} include biomedical texts, a set of Wikipedia articles, and datasets from DBpedia. Additionally, Braga et al. \cite{BCDM08} have evaluated the proposed optimization techniques using real data extracted by \url{www.conference-service.com}, \url{www.accuweather.com}, and \url{www.bookings.com}. Typically, when employing standard scientific flows, the datasets used are also fixed; however, in \cite{KGfgcs14} a wide-range of artificially created metadata have been used to cover more cases.

Finally, for many techniques, only small data flows comprising no more than 15 nodes were used, or the information with regards to the size of the flows could not be derived. In the latter case, this might be due to the fact that well-known algorithms have been used (e.g., k-means in \cite{CGD+15} and matrix-multi-plication in \cite{HBY13}) without explaining how these algorithms are internally translated to data flows. All experiments with workflows comprising hundreds of tasks used synthetic datasets.

\section{Discussion on findings}
\label{sec:discussion}
Data flow optimization is a research area with high potential for further improvements given the increasing role of data flows in modern data-driven applications. In this survey, we have listed more than thirty research proposals, most of which have been published after 2010.
In the previous sections, we mostly focused on the merits and the technical details of each proposal. They can lead to performance improvements, and more importantly, they have the potential  to lift the burden of manually fixing all implementation details from the data flow designers, which is a key motivation for automated optimization solutions. In this section, we complement any remarks made before with a list of additional observations, which may also serve as a description of directions for further research:

\begin{itemize}

\item In principle, the techniques described previously can serve as building block towards more holistic solutions. For instance, task ordering can, in principle, be combined with i) additional high-level mechanisms, such as task introduction, removal, merge, and decomposition; and ii) low-level mechanisms, such as engine configuration,  thus yielding added benefits. The main issue arising when mechanisms are combined is the increased complexity. An approach to mitigating the complexity is a two-phase approach, as commonly happens in database queries. Another issue is to determine which mechanism should first be explored. For some mechanisms, this is straight-forward, e.g., decomposition should precede task ordering and task removal should be placed afterwards. But, for mechanisms, such as configuration, this is unclear, e.g., whether it is beneficial to first configure low-level details before higher level ones remains an open issue.

\item In general, there is little work on low-complexity, holistic, and multi-objective solutions.
    Toward this direction, Simitsis et al. \cite{SWDC10} considers more than one objective and combines mechanisms at both high and low level execution plan details; for instance, both task ordering and engine configuration are addressed in the same technique. But clearly more work is needed here. In general, most of the techniques have been developed in isolation, each one typically assuming a specific setting and targeting a subset of optimization aspects. This and the lack of a common agreed benchmark makes it difficult to understand how exactly they compare to each other, the details of how the various proposals can be combined in a common framework and how they interplay.

\item There seems to be no common approach to evaluating the optimization proposals. Some proposals have not been adequately tested in terms of scalability, since they have considered only small graphs. In some data flow evaluations, workloads inspired from benchmarks such as TPC-DI/DS have been employed, but as most of the authors report as well, it is doubtful whether these benchmarks can completely capture all dimensions of the problem. There is a growing need for the development of systematic and broadly adopted techniques to evaluate optimization techniques for data flows.

\item A significant part of the techniques covered in this survey have not been incorporated in tools, nor have been exploited commercially.
Most of the optimization techniques described here, especially regarding the high level execution plan details, have not been implemented in real data flow systems apart from very few exceptions, as explained earlier. Hence, the full potential and practical value of the proposals have not been investigated in actual execution conditions, despite the fact that evaluation results thus far are shown to provide improvements by several orders of magnitude over non-optimized plans.

\item A plethora of objective functions and cost models have been investigated, which, to a large extent, they are compatible with each other, despite the fact that original proposals have examined them in isolation.
    However, it is unclear whether any of such cost models can capture aspects, such as the execution time of parallel data flows, which are very common nowadays, in a fairly accurate manner. A more sophisticated cost model  should take into account sequential, pipelined and partitioned execution in a unified manner, essentially combining the sum, bottleneck and critical path cost metrics.


\item Developing adaptive solutions that are capable of revising the flow execution plan on the fly is one important open issue, especially for online, continuous, and stream processing. Also, very few optimization techniques consider the cost of the graph edges. Not considering edge metadata does not
    reflect entirely real data flow execution in distributed settings, where the cost of transmitting data depends both on sender and receiver.

\item In this survey, we investigated  single flow optimizations. Optimizing multiple flows simultaneously, is another area requiring attention. An initial effort is described by Jovanovic et al. \cite{JRSA16}, which builds upon the task ordering solutions of \cite{SimVS05}.

\item There is early work on statistics collection \cite{HDP14,SBC06,CBK+2014,PJD+14}, but clearly, there is more to be done here given that without appropriate statistics, cost-based optimization becomes problematic and prone to significant errors.

\item On the other hand, a different school of thought advocates that in contrast to relational databases, automated optimization cannot help in practice in flow optimization due to flow complexity and increased difficulty in maintaining flow statistics, and developing accurate cost models. Based on that, there is a number of commercial flow execution engines (e.g., ETL tools) that instead of offering a flow optimizer they provide users with tips and best practices. No doubt, this is an interesting point, but we consider this category as out of the scope of this work.

\end{itemize}

Given the above observations and the trend in developing new solutions in the recent years, data flow optimization seems  to be technology in evolution rather than an area, where most significant problems have been resolved. Moreover, providing solutions to  all these problems is more likely to yield  significantly different and more powerful new approaches to data flow optimization, rather than delta improvements on existing solutions.

\section{Additional Issues in Data-centric Flow Optimization}
\label{sec:other}

Additional issues are  split into four parts. First, we describe optimizations enabled in current state-of-the-art parallel data flow systems, which, however, cannot cover arbitrary DAGs and tasks, and as such, have not been included in the previous sections. Next, we discuss techniques that, although they do not perform optimization in their own, they could, in principle, facilitate optimization. We provide a brief overview of optimization solutions for the WEP execution layer, complementing the discussion of existing scheduling techniques in Section~\ref{sec:rw}. We conclude with a brief note on implementing the optimization techniques into existing systems.

\subsection{Optimization In Massively Parallel Data Flow Systems}

A specific form of data flow systems are massively parallel processing (MPP) engines, such as Spark and Hadoop. These data flow systems can scale to a large number of computing nodes and are specifically tailored to big data management taking care of parallelism efficiency and fault tolerance issues. They accept their input in a declarative form (e.g., PigLatin~\cite{ORS+08}, Hive, SparkSQL), which is then automatically transformed into an executable DAG. Several optimizations take place during this transformation.

We broadly classify these optimizations in two categories. The first category comprises database-like optimizations, such as pushing filtering tasks as early as possible, choosing the join implementation, and using index tables, corresponding to \emph{task ordering} and \emph{implementation selection}, respectively. This can be regarded as a direct technology transfer from databases to parallel data flows and to date, these optimizations do not cover arbitrary user-defined transformations.

The second category is specific to  the parallel execution environment with a view to minimizing the amount of data read from disk, transmitted over the network, and being processed. For example, Spark groups pipelining tasks in larger jobs (called stages) to benefit from this type of parallelism. Also, it leverages cached data and columnar storage, performs compression, and reduces the amount of data transmitted during data shuffling through early partial aggregation, when this is possible. Grouping tasks into pipelining stages is a form of runtime scheduling.
Early partial aggregation can be deemed as a \emph{task introduction} technique. The other forms of of optimizations (leveraging cached data, columnar storage, and compression) can be deemed as specific forms of \emph{implementation selection}.
Flink is another system employing optimizations, but it has not yet incorporated all the (advanced) optimization proposals in its predecessor projects, as described in \cite{HPSR12,RHHLN15}. The proposal in \cite{BTR+14} is another example that proposes optimizations for a specific operator, namely \emph{ParFOR}.

We do not include these techniques in Tables~\ref{tab:opt1} and \ref{tab:opt2} because they apply to specific DAG instances and have not matured enough to benefit generic data flows including arbitrary tasks.

\subsection{Techniques Facilitating Data-centric Flow Optimization}
\label{sec:other-related}

Statistical metadata, such as cost per task invocation and selectivity, play a significant role in data flow optimization as discussed previously. \cite{CBK+2014,HDP14,PJD+14,SBC06} deal with statistics collection and modeling the execution cost of workflows; such issues are essential components in performing sophisticated flow optimization. \cite{VassiliadisSB09} analyze the properties of tasks, e.g., multiple-input vs single-input ones; such properties along with dependency constraint information complement statistics as the basis on top of which optimization solutions can be built.

Some techniques allow for choosing among multiple implementations of the same tasks using ontologies, rather than performing cost-based or heuristic optimization \cite{OODBM12}. In \cite{WBJM14}, improving the flow with the help of user interactions is discussed. Additionally, in \cite{OVDP11},  different scheduling strategies to account for data shipping between tasks are presented, without however proposing an optimization algorithm that takes decisions as to which strategy should be employed.

Apart from the optimizations described in Section~\ref{sec:classification}, the proposal in \cite{SWDC10} considers also the objective of data freshness. To this end, the proposal optimizes the activation time of ETL data flows, so that the changes in data sources are reflected on the state of a Data Warehouse within a time window. Nevertheless, this type of optimization objective leads to techniques that do not focus on optimizing the flow execution plan per se, which is the main topic of this survey.

For the evaluation of optimization proposals, benchmarks for evaluating techniques are proposed in \cite{VKTS07,SW12}. Finally, in \cite{HZH12,KMRHK09}, the significant role of correct parameter configuration in large-scale workflow execution is identified and relevant approaches are proposed. Proper tuning~of~the data flow execution environment is orthogonal and complementary to optimization of flow execution plan.

\subsection{On Scheduling Optimizations in Data-centric Flows}
\label{sec:sch}

In general, data flow execution engines tend to have built-in scheduling policies, which are not configured on a single flow basis. In principle, such policies can be extended to take into account the specific characteristics of data flows, where the placement of data and the transmission of data across tasks, represented by the DAG edges, requires special attention \cite{CD12}. For example, in \cite{KVS13}, a set of scheduling strategies for improving the performance through the minimization of memory consumption and the execution time of Extract-Transform-Load (ETL) workflows running on a single machine is proposed. As it is difficult to execute the data in pipeline in ETLs due to the blocking nature of some of the ETL tasks, the authors suggest splitting the workflow into several sub-flows and apply different scheduling policies if necessary. Finally, in \cite{JZS+13}, the placement of data management tasks is decided according to the memory availability of resources taking into account the trade-off between co-locating tasks and the increased memory consumption when running multiple tasks on the same physical computational node.

A large set of scheduling proposals target specific execution environments. For example, the technique in \cite{GWR11} targets shared resource
environments.  Proposals, such as \cite{CB14,ChenZ09,KMRHK09,RHRB13,SWH08,ZVZ15} are specific to grid and cloud data-centric flow scheduling. \cite{ABMR10} discusses optimal time schedules given a fixed allocation of tasks to engines, provided that the tasks belong to a linear workflow.

Also, a set of optimization algorithms for scheduling flows based on deadline and time constraints is analyzed  in \cite{ANE13,ANE12}. Another proposal of flow scheduling optimization is presented in \cite{PP2012} based on soft deadline rescheduling in order to deal with~the problem of  fault tolerance in flow executions. In \cite{CB14}, an optimization technique for minimizing the performance fluctuations that might occur by the resource diversity, which also considers deadlines, is proposed. Additionally, there is a set of scheduling techniques based on multi-objective optimization, e.g., \cite{FPF13}.

\subsection{On incorporation Optimization Techniques into Existing Systems}

Without loss of generality, there are two main types of describing the data flow execution plan in existing tools and prototypes: either in an appropriately formatted text file or using internal representations in the code. These two approaches are exemplified in systems, like the Pentaho Kettle, Spark, Taverna, and numerous others. In the former case, an optimization technique can be inserted as a component that processes this text file and produces a different execution plan. As an example, in Pentaho, each task and each graph edge are described as different XML elements in an XML document. Then, a technique that performs task reordering can consist of an independent programming module that parses the XML file and  modifies the edge elements. On the other hand, systems, such as Spark, transform the flow submitted by the user in a DAG, but without exposing a high level representation to the end user. The internal optimization component, called Catalyst, then performs modifications to the internal code structure that captures the executable DAG. Extending the optimizer to add new techniques, such as those described in this survey, requires using the Catalyst extensibility points. The second approach seems to require more effort from the developer and be more intrusive.

\section{Related Work}
\label{sec:rw}

To the best of our knowledge, there is no prior survey or overview article on data flow optimization; however, there are several surveys on related topics.

Related work falls into two categories: (i) surveys on generic DAG scheduling and on narrow-scope scheduling problems, which are also encountered in data flow optimization; and (ii) overviews of workflow systems.

DAG scheduling is a persisting topic in computing and has received a renewed attention due to the emergence of Grid and cloud infrastructures, which allow for the usage of remote computational resources. For such distributed settings, the proposals tend to refer to the WEP execution layer   and to focus on mapping computational tasks ignoring the data transfer between them, or assume a non-pipelined mode of execution that does not fit will into data-centric flow setting \cite{DA06}. A more recent survey of task mapping is presented in \cite{GDJ13}, which discusses techniques that assign tasks to resources for efficient execution in Grids under the demanding requirements and resource allocation constraints, such as the dependencies between the tasks, the resource reservation, and so on. In \cite{BCRS13}, an overview of the pipelined workflow time scheduling problem is presented, where the problem formulation targets streaming applications. In order to compare the effectiveness of the proposed optimization techniques, they present a taxonomy of workflow optimization techniques taking into account workflow characteristics, such as the structure of flow (i.e., linear, fork, tree-shaped DAGs), the computation requirements, the size of data to be transferred between tasks, the parallel or sequential task execution mode, and the possibility of executing task replicas. Additionally, the taxonomy takes into consideration a performance model that describes whether the optimization aims to a single or multiple objectives, such as throughput, latency, reliability, and so on. However, in data-centric flows, tasks are activated upon receipt of input data and not as a result of an activation message from a controller, as assumed in \cite{BCRS13}. None of the surveys above provides a systematic study of the optimizations at the WEP generation layer.

The second class of related work deals with a broader-scope presentation of workflow systems. The survey in \cite{DGST09} aims to present a taxonomy of the workflow system features and capabilities to allow end users to take the best option for each application. Specifically, the taxonomy is inspired by the workflow lifecycle and categorizes the workflow systems according to the lifecycle phase they are capable of supporting. However, the optimizations considered suffer from the same limitations as those in \cite{DA06}. Similarly, in \cite{BH07}, an evaluation of the current workflow technology is also described, considering both scientific and business workflow frameworks. The control and data flow mechanisms and capabilities of workflow systems both for e-science, e.g., Taverna and Triana, and business processes, e.g., YAWL and BPEL-based engines, are discussed in \cite{CG08}. \cite{VSRM08} discusses how leading commercial tools in the data analysis market handle SQL statements, as a means to perform data management tasks within workflows.  Liu et al. \cite{LPVM15} focus on scientific workflows, which are an essential part of data flows, but does not delve into the details of optimization. Finally, Jovanovic et al. \cite{Jovanovic16} present a survey that aims to present the challenges of modern data flows through different data flow scenarios. Additionally, related data flow optimization techniques are summarized, but not surveyed, in order to underline the importance of low data latency in Business Intelligence (BI) processes, while an architecture of next generation BI systems that manage the complexity of modern data flows in such systems is proposed.

Modeling and processing ETL workflows \cite{V09} focuses on the detailed description of conceptual and logical modeling of ETLs. Conceptual modeling refers to the initial design of ETL processes by using UML diagrams, while the logical modeling refers to the design of ETL processes taking into account required constraints. This survey discusses the generic problems in ETL data flows, including optimization issues in minimizing the execution time of an ETL workflow and the resumption in case of failures during the processing of large amount of data.

Data flow optimization bears also similarities with query optimization over Web Services (WSs) \cite{TsamouraGM11}, especially when the valid orderings of the calls to the WSs are subject to dependency constraints. This survey includes all the WSs related techniques that can also be applied to data flows.

Part of the optimizations covered in this survey can be deemed as generalizations of the corresponding techniques in database queries. An example is the correspondence between pushing selections down in the query plan and moving filtering tasks as close to data source as possible \cite{Bohm2011}. Comprehensive surveys on database query optimization are in \cite{Cha98,Ioa96}, whereas lists of semantic equivalence rules between expressions of relational operators that provide the basis for query optimization can be found in classical database textbooks (e.g., \cite{dbsystems}). However, as discussed in the introduction, there are essential differences between database queries and data flows, which cannot be described as expressions over a limited set of elementary operations. At a higher level, data flow optimization  covers more mechanisms (e.g., task decomposition and engine selection) and a broader setting with regards to the criteria considered and the metadata required.

Nevertheless, it is arguable that data flow task ordering bears similarities to optimization of database queries containing user-defined functions (UDFs) (or, expensive predicates), as reported in \cite{CS99,Hel98}. This similarity is based on the intrinsic correspondence between UDFs and data flow tasks, but there are two main differences. First, the dependency constraints considered in \cite{CS99,Hel98} refer to pairs of a join and a UDF, rather than between UDFs. As such, when joins are removed and only UDFs are considered, the techniques described in these proposals are reduced to unconstrained filter ordering. Second, the straightforward extensions to the proposals \cite{CS99,Hel98} are already covered and improved by solutions targeting data flow task ordering explicitly as discussed in Section \ref{sec:ordering}.

\section{Summary}
\label{sec:summary}

This survey covers an emerging area in data management, namely  optimization techniques that modify a data-centric workflow execution plan prior to its execution in an automated manner. The survey first provides a taxonomy of the main dimensions characterizing each optimization proposal. These dimensions cover a broad range, from the mechanism utilized to enhance execution plans to the distribution of the setting and the environment for which the solution is initially proposed. Then, we present the details of the existing proposals, divided into eight groups, one for each of the identified optimization mechanisms. Next, we present the evaluation approaches, focusing on aspects, such as the type of workflows and data used during experiments. We complete this survey with a discussion of the main findings, while also, for completeness, we briefly  present tangential issues, such as optimizations in massively parallel data flow systems and optimized workflow scheduling.

\bibliographystyle{model1-num-names}
\bibliography{csurvey-bibfile}







\appendix

\end{document}